\documentclass[runningheads]{llncs}
\usepackage{graphicx}
\usepackage{amsfonts}
\usepackage{amsmath}
\usepackage{amssymb}
\begin{document}
\title{Comparative Analysis of Deterministic and Nondeterministic Decision Trees for Decision Tables from Closed Classes}
\titlerunning{Comparative Analysis of Deterministic and Nondeterministic Decision Trees}
%
\author{
Azimkhon Ostonov \and Mikhail Moshkov
}
\authorrunning{Azimkhon Ostonov \and Mikhail Moshkov}
%
%
\institute{
Computer, Electrical and Mathematical Sciences \& Engineering Division\\
and Computational Bioscience Research Center\\
King Abdullah University of Science and Technology (KAUST)\\
Thuwal 23955-6900, Saudi Arabia\\
\email{\{azimkhon.ostonov,mikhail.moshkov\}@kaust.edu.sa}
}
\maketitle              
\begin{abstract}
In this paper, we consider classes of decision tables
with many-valued decisions closed under operations of removal of columns,
changing of decisions, permutation of columns, and duplication of co\-lumns.
We study relationships among three parameters of these tables: the
complexity of a decision table (if we consider the depth of decision trees,
then the complexity of a decision table is the number of columns in it), the
minimum complexity of a deterministic decision tree, and the minimum
complexity of a nondeterministic decision tree. We consider rough
classification of functions characterizing relationships and enumerate all possible seven types of the relationships.

\end{abstract}
\section{Introduction} \label{S1}
In this paper, we consider closed classes of decision tables with
many-valued decisions and study relationships among three parameters of
these tables: the complexity of a decision table (if we consider the depth of
decision trees, then the complexity of a decision table is the number of
columns in it), the minimum complexity of a deterministic decision tree, and
the minimum complexity of a nondeterministic decision tree.

A decision table with many-valued decisions is a rectangular table in which
columns are labeled with attributes, rows are pairwise different and each
row is labeled with a nonempty finite set of decisions. Rows are interpreted
as tuples of values of the attributes. For a given row, it is required to
find a decision from the set of decisions attached to the row. To this end,
we can use the following queries: we can choose an attribute and ask what is
the value of this attribute in the considered row. We study two types of
algorithms based on these queries: deterministic and nondeterministic
decision trees. One can interpret nondeterministic decision trees for a
decision table as a way to represent an arbitrary system of true decision
rules for this table that cover all rows. We consider in some sense
arbitrary complexity measures that characterize the time complexity of decision
trees. Among them, we distinguish so-called limited complexity measures, for
example, the depth of decision trees.

Decision tables with many-valued decisions often appear in data analysis,
where they are known as multi-label decision tables \cite%
{Boutell04,Vens08,Zhou12}. Moreover, decision tables with many-valued
decisions are common in such areas as combinatorial optimization,
computational geometry, and fault diagnosis, where they are used to represent
and explore problems \cite{book20,MoshkovZ11}.

Decision trees \cite%
{AbouEishaACHM19,book20,BreimanFOS84,Moshkov05,Moshkov20,Quinlan93,RokachM07}
and decision rule systems  \cite%
{BorosHIK97,BorosHIKMM00,ChikalovLLMNSZ13,FurnkranzGL12,MPZ08,MoshkovZ11,Pawlak91,PawlakS07}
are widely used as classifiers, as a means for knowledge representation, and
as algorithms for solving various problems of combinatorial optimization,
fault diagnosis, etc. Decision trees and rules are among the most
interpretable models in data analysis \cite{Molnar22}.

The depth of deterministic and nondeterministic decision trees for
computation Boolean functions (variables of a function are considered as
attributes) was studied quite intensively \cite%
{BlumI87,HartmanisH87,Moshkov95,Tardos89}. Note that the minimum depth of a
nondeterministic decision tree for a Boolean function is equal to its
certificate complexity \cite{BuhrmanW02}.

We study classes of decision tables with many-valued decisions closed under
four operations: removal of columns, changing of decisions, permutation of
columns, and duplication of columns. The most natural examples of such
classes are closed classes of decision tables generated by information
systems \cite{Pawlak81}. An information system consists of a set of objects
(universe) and a set of attributes (functions) defined on the universe and
with values from a finite set. A problem over an information system is
specified by a finite number of attributes that divide the universe into
nonempty domains in which these attributes have fixed values. A nonempty
finite set of decisions is attached to each domain. For a given object from
the universe, it is required to find a decision from the set attached to the
domain containing this object.

A decision table with many-valued decisions
corresponds to this problem in a natural way: columns of this table are
labeled with the considered attributes, rows correspond to domains and are
labeled with sets of decisions attached to domains. The set of decision
tables corresponding to problems over an information system forms a closed class
generated by this system. Note that the family of all closed classes is
essentially wider than the family of closed classes generated by information
systems. In particular, the union of two closed classes generated by two
information systems is a closed class. However, generally, there is no an
information system that generates this class.

Various classes of objects that are closed under different operations are
intensively studied. Among them, in particular, are classes of Boolean
functions closed under the operation of superposition \cite{Post41},
minor-closed classes of graphs \cite{Robertson04}, classes of read-once
Boolean functions closed under removal of variables and renaming of
variables \cite{Lozin21}, languages closed under taking factors \cite%
{Atminas17}, etc. Decision tables represent an interesting mathematical
object deserving mathematical research, in particular, the study of closed
classes of decision tables.

This paper continues the study of closed classes of decision tables that started
by work \cite{Moshkov89} and frozen for various reasons for many years. In
\cite{Moshkov89}, we studied the dependence of the minimum depth of
deterministic decision trees and the depth of deterministic decision trees
constructed by a greedy algorithm on the number of attributes (columns) for
conventional decision tables from classes closed under operations of removal
of columns and changing of decisions.

In the present paper, we study so-called t-pairs $(\mathcal{C},\psi )$,
where $\mathcal{C}$ is a class of decision tables closed under the
considered four operations and $\psi $ is a complexity measure for this
class. The t-pair is called limited if $\psi $ is a limited complexity
measure. For any decision table $T\in \mathcal{C}$, we have three parameters:

\begin{itemize}
\item $\psi ^{i}(T)$ -- the complexity of the decision table $T$. This
parameter is equal to the complexity of a deterministic decision tree for
the table $T$, which sequentially computes values of all attributes attached
to columns of $T$.

\item $\psi ^{d}(T)$ -- the minimum complexity of a deterministic decision
tree for the table $T$.

\item $\psi ^{a}(T)$ -- the minimum complexity of a nondeterministic
decision tree for the table $T$.
\end{itemize}

We investigate the relationships between any two such parameters for
decision tables from $\mathcal{C}$. Let us consider, for example, the
parameters $\psi ^{i}(T)$ and $\psi ^{d}(T)$. Let $n\in \mathbb{N}$. We will
study relations of the kind $\psi ^{i}(T)\leq n\Rightarrow \psi ^{d}(T)\leq
u $, which are true for any table $T\in \mathcal{C}$. The minimum value of $u
$ is the most interesting for us. This value (if exists) is equal to
\[
\mathcal{U}_{\mathcal{C}\psi }^{di}(n)=\max \left\{ \psi ^{d}(T):T\in
\mathcal{C},\psi ^{i}(T)\leq n\right\}.
\]%
We will also study relations of the kind $\psi ^{i}(T)\geq n\Rightarrow \psi
^{d}(T)\geq l$. In this case, the maximum value of $l$ is the most
interesting for us. This value (if exists) is equal to
\[
\mathcal{L}_{\mathcal{C}\psi }^{di}(n)=\min \left\{ \psi ^{d}(T):T\in
\mathcal{C},\psi ^{i}(T)\geq n\right\}.
\]%
The two functions $\mathcal{U}_{\mathcal{C}\psi }^{di}$ and $\mathcal{L}_{%
\mathcal{C}\psi }^{di}$ describe how the behavior of the parameter $\psi
^{d}(T)$ depends on the behavior of the parameter $\psi ^{i}(T)$ for tables
from $\mathcal{C}$.

There are $18$ similar functions for all ordered pairs of parameters $\psi
^{i}(T)$, $\psi ^{d}(T)$, and $\psi ^{a}(T)$. These $18$ functions well
describe the relationships among the considered parameters. It would be very
interesting to point out $18$-tuples of these functions for all t-pairs and
all limited t-pairs. But this is a very difficult problem.

In this paper,
instead of functions, we will study types of functions. With any partial
function $f:\mathbb{N\rightarrow N}$, we will associate its type from the
set $\{\alpha ,\beta ,\gamma ,\delta ,\epsilon \}$. For example, if the
function $f$ has an infinite domain, and it is bounded from above, then its
type is equal to $\alpha $. If the function $f$ has an infinite domain, is not
bounded from above, and the inequality $f(n)\geq n$ holds for a finite
number of $n\in \mathbb{N}$, then its type is equal to $\beta $, etc. Thus,
we will enumerate $18$-tuples of types of functions. These tuples will be
represented in tables called the types of t-pairs. We will prove that there
are only seven realizable types of t-pairs and only five realizable types of
limited t-pairs.

First, we will study $9$-tuples of types of functions  $\mathcal{U}_{%
\mathcal{C}\psi }^{bc}$, $b,c\in \{i,d,a\}$. These tuples will be
represented in tables called upper types of t-pairs. We will enumerate all
realizable upper types of t-pairs and limited t-pairs. After that, we will
extend the results obtained for upper types of t-pairs to the case of types
of t-pairs. We will also define the notion of a union of two t-pairs
and study the upper type of the resulting t-pair depending on the upper
types of the initial t-pairs.

This paper is based on the work \cite{Moshkov05a} in which similar results
were obtained for classes of problems over information systems. We
generalized proofs from \cite{Moshkov05a} to the case of decision tables
from closed classes and use some results from this paper to prove the
existence of t-pairs and limited t-pairs with given upper types.

The paper consists of eight sections. In Sect. \ref{S2}, basic definitions are
considered. In Sect. \ref{S3}, we provide the main results related to types of t-pairs
and limited t-pairs. In Sects. \ref{S4}-\ref{S6}, we study upper types of t-pairs and
limited t-pairs. Section \ref{S7} contains proofs of the main results and Sect. \ref{S8} --
short conclusions.
\section{Basic Definitions}  \label{S2}
\subsection{Decision Tables and Closed Classes}
Let $\mathbb{N} = \{0,1,2,\ldots\}$ be the set of nonnegative integers. For any $k\in \mathbb{N}\setminus\{0, 1\}$, let $E_{k} = \{0,1,\ldots,k-1\}$. The set of nonempty finite subsets of the set $\mathbb{N}$ will be denoted by $\mathcal{P}(\mathbb{N})$. 
Let $F$ be a nonempty set of \textit{attributes} (really, names of attributes). 

\begin{definition}
We now define the set of decision tables $\mathcal{M}_{k}(F)$. An arbitrary \textit{decision table} $T$ from this set is a rectangular table with $n\in \mathbb{N}\setminus\{0\}$ columns labeled with attributes $f_{1}, \ldots, f_{n}\in F$, where any two columns labeled with the same attribute are equal. The rows of this table are pairwise different and are filled in with numbers from $E_{k}$. Each row is interpreted as a tuple of values of attributes $f_{1}, \ldots, f_{n}$. For each row in the table, a set from $\mathcal{P}(\mathbb{N})$ is attached, which is interpreted as a set of decisions for this row.
\end{definition}

\begin{example}
\label{example1}
Three decision tables $T_{1}$, $T_{2}$, and $T_{3}$ from the set $\mathcal{M}_{2}(F_{0})$, where $F_{0}=\{f_{1}, f_{2}, f_{3}\}$, are shown in Fig. \ref{fig:fig1}.
\end{example}
\begin{figure}
\begin{minipage}[c]{0.3\textwidth}
\begin{center}
$T_{1}=$
\begin{tabular}{ |cc|c| }
 \hline
 $f_{1}$ & $f_{2}$ & \\
  \hline
 0 & 0  & $\{1\}$ \\
 1 & 0  & $\{2, 3\}$\\
 0 & 1  & $\{2\}$ \\
 1 & 1 & $\{4\}$\\
 \hline
\end{tabular}
\end{center}
\end{minipage}
\begin{minipage}[c]{0.3\textwidth}
\begin{center}
$T_{2}=$
\begin{tabular}{ |ccc|c| }
 \hline
 $f_{1}$ & $f_{2}$ & $f_{3}$ & \\
  \hline
 1 & 0 & 0  & $\{1, 2\}$ \\
 0 & 1  & 0  & $\{1, 3\}$ \\
 0 & 0  & 1  & $\{4\}$\\
 0 & 0  & 0  & $\{1,2,3\}$\\
 \hline
\end{tabular}
\end{center}
\end{minipage}
\begin{minipage}[c]{0.3\textwidth}
\begin{center}
$T_{3}=$
\begin{tabular}{ |ccc|c| }
 \hline
 $f_{1}$ & $f_{1}$ & $f_{3}$ & \\
  \hline
 0 & 0 & 0  & $\{1, 3\}$ \\
  1 & 1  & 0  & $\{1\}$\\
 0 & 0  & 1  & $\{2\}$ \\
 1 & 1  & 1  & $\{1,2\}$\\
 \hline
\end{tabular}
\end{center}
\end{minipage}
\caption{Decision tables $T_{1}$, $T_{2}$, and $T_{3}$}
\label{fig:fig1}
\end{figure}

We correspond to the table $T$ the following \textit{problem}: for a given row of $T$, we should recognize a decision from the set of decisions attached to this row. To this end, we can use queries about the values of attributes for this row.

We denote by $At(T)$ the set $\{f_{1},\ldots,f_{n}\}$ of attributes attached to the columns of $T$. By $\Pi(T)$, we denote the intersection of the sets of decisions attached to the rows of $T$, and by $\Delta(T)$, we denote the set of rows of the table $T$. Decisions from $\Pi(T)$ are called \textit{common decisions} for $T$. The table $T$ will be called \textit{degenerate} if $\Delta(T)=\varnothing$ or $\Pi(T)\neq\varnothing$. We denote by $\mathcal{M}_{k}^{c}(F)$ the set of degenerate decision tables from $\mathcal{M}_{k}(F)$.

\begin{example}
\label{example2}
Two degenerate decision tables $D_{1}$ and $D_{2}$ are shown in Fig. \ref{fig:fig2}.
\end{example}
\begin{figure}
\begin{minipage}[c]{0.4\textwidth}
\begin{center}
$D_{1}=$
\begin{tabular}{ |cc|c| }
 \hline
 $f_{1}$ & $f_{2}$ & \\
  \hline
\end{tabular}
\end{center}
\end{minipage}
\begin{minipage}[c]{0.4\textwidth}
\begin{center}
$D_{2}=$
\begin{tabular}{ |ccc|c| }
 \hline
 $f_{1}$ & $f_{2}$ & $f_{3}$ & \\
  \hline
 1 & 0 & 0  & $\{1, 2\}$ \\
 0 & 1  & 0  & $\{1, 3\}$ \\
 0 & 0  & 0  & $\{1,2,3\}$\\
 \hline
\end{tabular}
\end{center}
\end{minipage}
\caption{Degenerate decision tables $D_{1}$ and $D_{2}$}
\label{fig:fig2}
\end{figure}

\begin{definition}
A\textit{ subtable} of the table $T$ is a table obtained from $T$ by removal of some of its rows. Let $\Theta(T)=\{(f, \delta): f\in At(T), \delta \in E_{k}\}$ and $\Theta^{*}(T)$ be the set of all finite words in the alphabet $\Theta(T)$ including the empty word $\lambda$. Let $\alpha\in \Theta^{*}(T)$. We now define a subtable $T\alpha$ of the table $T$. If $\alpha=\lambda$, then $T\alpha=T$. Let $\alpha=(f_{i_{1}}, \delta_{1})\cdots(f_{i_{m}}, \delta_{m})$. Then $T\alpha$ consists of all rows of $T$ that in the intersection with columns $f_{i_{1}},\ldots,f_{i_{m}}$ have values $\delta_{1},\ldots,\delta_{m}$, respectively.
\end{definition}

\begin{example}
\label{example3}
Two subtables of the tables $T_{1}$ and $T_{2}$ (depicted in Fig. \ref{fig:fig1}) are shown in Fig. \ref{fig:fig3}.
\end{example}
\begin{figure}
    \begin{minipage}[c]{0.4\textwidth}
\begin{center}
$T_{1}(f_{1}, 1)=$
\begin{tabular}{ |cc|c| }
 \hline
 $f_{1}$ & $f_{2}$ & \\
  \hline
 1 & 0  & $\{2, 3\}$\\
 1 & 1 & $\{4\}$\\
 \hline
\end{tabular}
\end{center}
\end{minipage}
\begin{minipage}[c]{0.6\textwidth}
\begin{center}
$T_{2}(f_{1}, 0)(f_{2}, 0)(f_{3}, 0)=$
\begin{tabular}{ |ccc|c| }
 \hline
 $f_{1}$ & $f_{2}$ & $f_{3}$ & \\
  \hline
 0 & 0  & 0  & $\{1,2,3\}$\\
 \hline
\end{tabular}
\end{center}
\end{minipage}
    \caption{Subtables $T_{1}(f_{1}, 1)$ and $T_{2}(f_{1}, 0)(f_{2}, 0)(f_{3}, 0)$ of tables $T_{1}$ and $T_{2}$ shown in Fig. \ref{fig:fig1}}
    \label{fig:fig3}
\end{figure}


We now define four operations on the set $\mathcal{M}_{k}(F)$ of decision tables:

    \begin{definition}
    \textit{ Removal of columns.} We can remove an arbitrary column in a table $T$ with at least two columns. As a result, the obtained table can have groups of equal rows. We keep only the first row in each such group.
    \end{definition}
   \begin{definition}\textit{Changing of decisions.} In a given table $T$, we can change in an arbitrary way sets of decisions attached to rows.
    \end{definition}
   \begin{definition}\textit{Permutation of columns.} We can swap any two columns in a table $T$, including the attached attribute names.
    \end{definition}
  \begin{definition}\textit{Duplication of columns.} For any column in a table $T$, we can add its duplicate next to that column.
    \end{definition}

\begin{example}
\label{example4}
Decision tables $T_{1}^{\prime}$,$T_{2}^{\prime}$,$T_{1}^{\prime\prime}$, and $T_{2}^{\prime\prime}$ depicted in Fig. \ref{fig:fig4} are obtained from decision tables $T_{1}$ and $T_{2}$ shown in Fig. \ref{fig:fig1} by operations of changing the decisions, removal of columns, permutation of columns, and duplication of columns, respectively.
\end{example}
\begin{figure}
\begin{minipage}[c]{0.21\textwidth}
\begin{center}
$T_{1}^{\prime}=$
\begin{tabular}{ |cc|c| }
 \hline
 $f_{1}$ & $f_{2}$ & \\
  \hline
 0 & 0  & $\{1, 4\}$ \\
  1 & 0  & $\{2, 3\}$\\
 0 & 1  & $\{3\}$ \\
 1 & 1 & $\{4\}$\\
 \hline
\end{tabular}
\end{center}
\end{minipage}
\begin{minipage}[c]{0.17\textwidth}
\begin{center}
$T_{2}^{\prime}=$
\begin{tabular}{ |c|c| }
 \hline
 $f_{1}$ & \\
  \hline
 1 & $\{1, 2\}$ \\
 0 & $\{1, 3\}$ \\
 \hline
\end{tabular}
\end{center}
\end{minipage}
\begin{minipage}[c]{0.21\textwidth}
\begin{center}
$T_{1}^{\prime\prime}=$
\begin{tabular}{ |cc|c| }
 \hline
 $f_{2}$ & $f_{1}$ & \\
  \hline
 0 & 0  & $\{1\}$ \\
  0 & 1  & $\{2, 3\}$\\
 1 & 0  & $\{2\}$ \\
 1 & 1 & $\{4\}$\\
 \hline
\end{tabular}
\end{center}
\end{minipage}
\begin{minipage}[c]{0.31\textwidth}
\begin{center}
$T_{2}^{\prime\prime}=$
\begin{tabular}{ |cccc|c| }
 \hline
 $f_{1}$ & $f_{2}$ & $f_{2}$ & $f_{3}$ & \\
  \hline
 1 & 0 & 0 & 0  & $\{1, 2\}$ \\
 0 & 1 & 1 & 0  & $\{1, 3\}$ \\
 0 & 0 & 0 & 1  & $\{4\}$\\
 0 & 0 & 0 & 0  & $\{1,2,3\}$\\
 \hline
\end{tabular}
\end{center}
\end{minipage}
    \caption{Decision tables $T_{1}^{\prime}$,$T_{2}^{\prime}$,$T_{1}^{\prime\prime}$, and $T_{2}^{\prime\prime}$ obtained from tables $T_{1}$ and $T_{2}$ shown in Fig. \ref{fig:fig1} by operations of changing the decisions, removal of columns, permutation of columns and duplication of columns, respectively}
    \label{fig:fig4}
\end{figure}
\begin{definition}
    Let $T\in \mathcal{M}_{k}(F)$. The closure of the table $T$ is a set, which contains all tables that can be obtained from $T$ by the operations of removal of columns, changing of decisions, permutation of columns, and duplication of columns and only such tables. We denote the closure of the table $T$ by $[T]$. It is clear that $T\in [T]$.
\end{definition}
\begin{definition}
Let $\mathcal{C}\subseteq\mathcal{M}_{k}(F)$. The closure $[\mathcal{C}]$ of the set $\mathcal{C}$ is defined in the following way: $[\mathcal{C}]=\bigcup_{T\in \mathcal{C}} [T]$. We will say that $\mathcal{C}$ is a closed class if $\mathcal{C}=[\mathcal{C}]$. In particular, the empty set of tables is a closed class.
\end{definition}

\begin{example}
\label{example5}
We now consider a closed class $\mathcal{C}_{0}$ of decision tables from the set $\mathcal{M}_{2}(\{f_{1}, f_{2}\})$, which is equal to $[Q]$, where the decision table $Q$ is depicted in Fig. \ref{fig:fig5_2}. The closed class $\mathcal{C}_{0}$ contains all tables depicted in Fig. \ref{fig:fig5} and all tables that can be obtained from them by operations of duplication of columns and permutation of columns. 
\end{example}
\begin{figure}
\begin{minipage}[c]{1.0\textwidth}
\begin{center}
$Q=$
\begin{tabular}{ |cc|c| }
 \hline
 $f_{1}$ & $f_{2}$ & \\
  \hline
 1 & 0 & $\{1\}$ \\
 0 & 1  & $\{2\}$ \\
 0 & 0  & $\{3\}$ \\
 \hline
\end{tabular}
\end{center}
\end{minipage}
\caption{Decision table $Q$}
\label{fig:fig5_2}
\end{figure}

\begin{figure}
\begin{minipage}[c]{0.3\textwidth}
\begin{center}
$Q_{1}=$
\begin{tabular}{ |cc|c| }
 \hline
 $f_{1}$ & $f_{2}$ & \\
  \hline
 1 & 0 & $d_{1}$ \\
 0 & 1  & $d_{2}$ \\
 0 & 0  & $d_{3}$ \\
 \hline
\end{tabular}
\end{center}
\end{minipage}
\begin{minipage}[c]{0.3\textwidth}
\begin{center}
$Q_{2}=$
\begin{tabular}{ |c|c| }
 \hline
 $f_{1}$ & \\
  \hline
 1 & $d_{4}$ \\
 0 & $d_{5}$ \\
 \hline
\end{tabular}
\end{center}
\end{minipage}
\begin{minipage}[c]{0.3\textwidth}
\begin{center}
$Q_{3}=$
\begin{tabular}{ |c|c| }
 \hline
 $f_{2}$ & \\
  \hline
 0 & $d_{6}$ \\
 1 & $d_{7}$ \\
 \hline
\end{tabular}
\end{center}
\end{minipage}
    \caption{Decision tables from closed class $\mathcal{C}_{0}$, where $d_{1},\ldots,d_{7}\in \mathcal{P}(\mathbb{N})$}
    \label{fig:fig5}
\end{figure}

If $\mathcal{C}_{1}$ and $\mathcal{C}_{2}$ are closed classes belonging to $\mathcal{M}_{k}(F)$, then $\mathcal{C}_{1}\cup\mathcal{C}_{2}$ is also a closed class. We can consider closed classes $\mathcal{C}_{1}$ and $\mathcal{C}_{2}$ belonging to different sets of decision tables. Let $\mathcal{C}_{1}\subseteq\mathcal{M}_{k_{1}}(F_{1})$ and $\mathcal{C}_{2}\subseteq\mathcal{M}_{k_{2}}(F_{2})$. Then $\mathcal{C}_{1}\cup \mathcal{C}_{2}$ is a closed class and $\mathcal{C}_{1}\cup\mathcal{C}_{2}\subseteq\mathcal{M}_{\max(k_{1}, k_{2})}(F_{1}\cup F_{2})$.

\subsection{Deterministic and Nondeterministic Decision Trees}
A \textit{finite directed tree with the root} is a finite directed tree in which exactly one node has no entering edges. This node is called the \textit{root}. Nodes of the tree, which have no outgoing edges are called \textit{terminal} nodes. Nodes that are neither the root nor the terminal are called \textit{worker} nodes. A \textit{complete path} in a finite directed tree with the root is any sequence of nodes and edges starting from the root node and ending with a terminal node $\xi = v_{0},d_{0},\ldots,v_{m},d_{m},v_{m+1}$, where $d_{i}$ is the edge outgoing from the node $v_{i}$ and entering the node $v_{i+1}, i=0,\ldots,m$.

\begin{definition}
A decision tree over the set of decision tables $\mathcal{M}_{k}(F)$ is a labeled finite directed tree with the root with at least two nodes (the root and a terminal node) possessing the following properties:
\begin{itemize}
    \item[$\bullet$] The root and the edges outgoing from the root are not labeled.
    \item[$\bullet$] Each worker node is labeled with an attribute from the set $F$.
    \item[$\bullet$] Each edge outgoing from a worker node is labeled with a number from $E_{k}$.
    \item[$\bullet$] Each terminal node is labeled with a number from $\mathbb{N}$.
\end{itemize}
\end{definition}
We denote by $\mathcal{T}_{k}(F)$ the set of decision trees over the set of decision tables $\mathcal{M}_{k}(F)$.
\begin{definition}
A decision tree from $\mathcal{T}_{k}(F)$ is called \textit{deterministic} if it satisfies the following conditions:
\begin{itemize}
    \item[$\bullet$] Exactly one edge leaves the root.
    \item[$\bullet$] Edges outgoing from each worker node are labeled with pairwise different numbers.
\end{itemize}
\end{definition}

Let $\Gamma$ be a decision tree from $\mathcal{T}_{k}(F)$. Denote by $At(\Gamma)$ the set of attributes attached to worker nodes of $\Gamma$. Set $\Theta(\Gamma)=\{(f, \delta):f\in At(\Gamma), \delta\in E_{k}\}$. Denote by $\Theta^{*}(\Gamma)$ the set of all finite words in the alphabet $\Theta(\Gamma)$ including the empty word $\lambda$. We correspond to an arbitrary complete path $\xi=v_{0}, d_{0}, \ldots, v_{m}, d_{m}, v_{m+1}$ in $\Gamma$, a word $\pi(\xi)$. If $m=0$, then $\pi(\xi)=\lambda$. Let $m>0$ and, for $i=1,\ldots,m$, the node $v_{i}$ be labeled with an attribute $f_{j_{i}}$ and the edge $d_{i}$ be labeled with the number $\delta_{i}$. Then $\pi(\xi)=(f_{j_{1}},\delta_{1})\cdots(f_{j_{m}},\delta_{m})$. We denote by $\tau(\xi)$ the number attached to the terminal node of the path $\xi$. We denote by $Path(\Gamma)$ the set of complete paths in the tree $\Gamma$.
\begin{definition}
Let $T\in \mathcal{M}_{k}(F)$. A \textit{nondeterministic decision tree} for the table $T$ is a decision tree $\Gamma$ over $\mathcal{M}_{k}(F)$ satisfying the following conditions:
\begin{itemize}
    \item $At(\Gamma)\subseteq At(T).$
    \item $\bigcup_{\xi \in Path(\Gamma)} \Delta(T\pi(\xi))=\Delta(T).$
    \item For any row $r\in \Delta(T)$ and any complete path $\xi\in Path(\Gamma)$, if $r\in\Delta(T\pi(\xi))$, then $\tau(\xi)$ belongs to the set of decisions attached to the row $r$.
\end{itemize}
\end{definition}

\begin{figure}
  \centering
  \includegraphics[width=0.8\columnwidth]{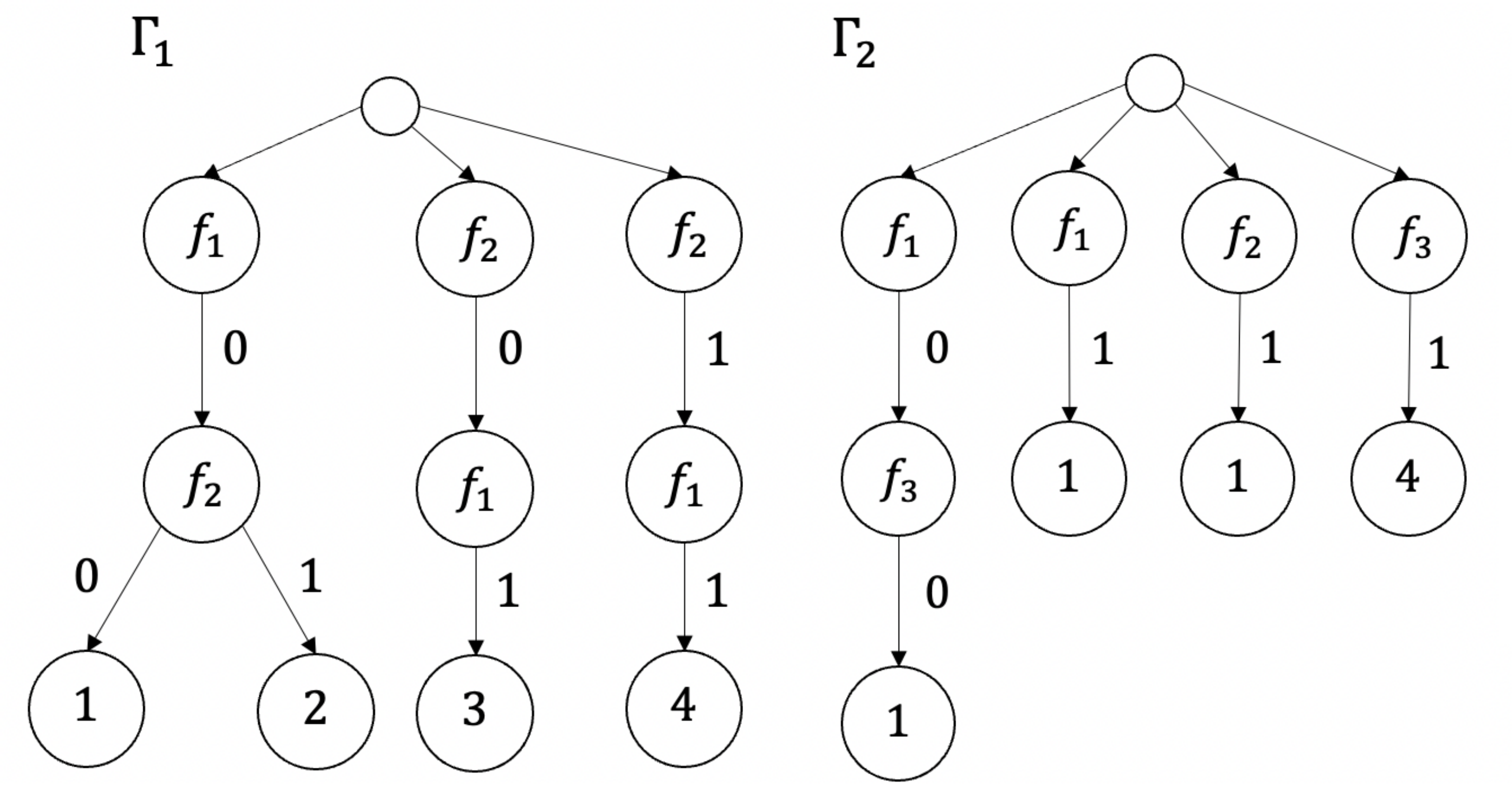}
  \caption{Nondeterministic decision trees $\Gamma_{1}$ and $\Gamma_{2}$ for decision tables $T_{1}$ and $T_{2}$ depicted in Fig. \ref{fig:fig1}}
  \label{fig:fig6}
\end{figure}

\begin{example}
\label{example6}
Nondeterministic decision trees $\Gamma_{1}$ and $\Gamma_{2}$ for decision tables $T_{1}$ and $T_{2}$ shown in Fig. \ref{fig:fig1} are depicted in Fig. \ref{fig:fig6}.
\end{example}

\begin{definition}
A \textit{deterministic decision tree} for the table $T$ is a deterministic decision tree over $\mathcal{M}_{k}(F)$, which is a nondeterministic decision tree for the table $T$.
\end{definition}

\begin{figure}
  \centering
  \includegraphics[width=0.8\columnwidth]{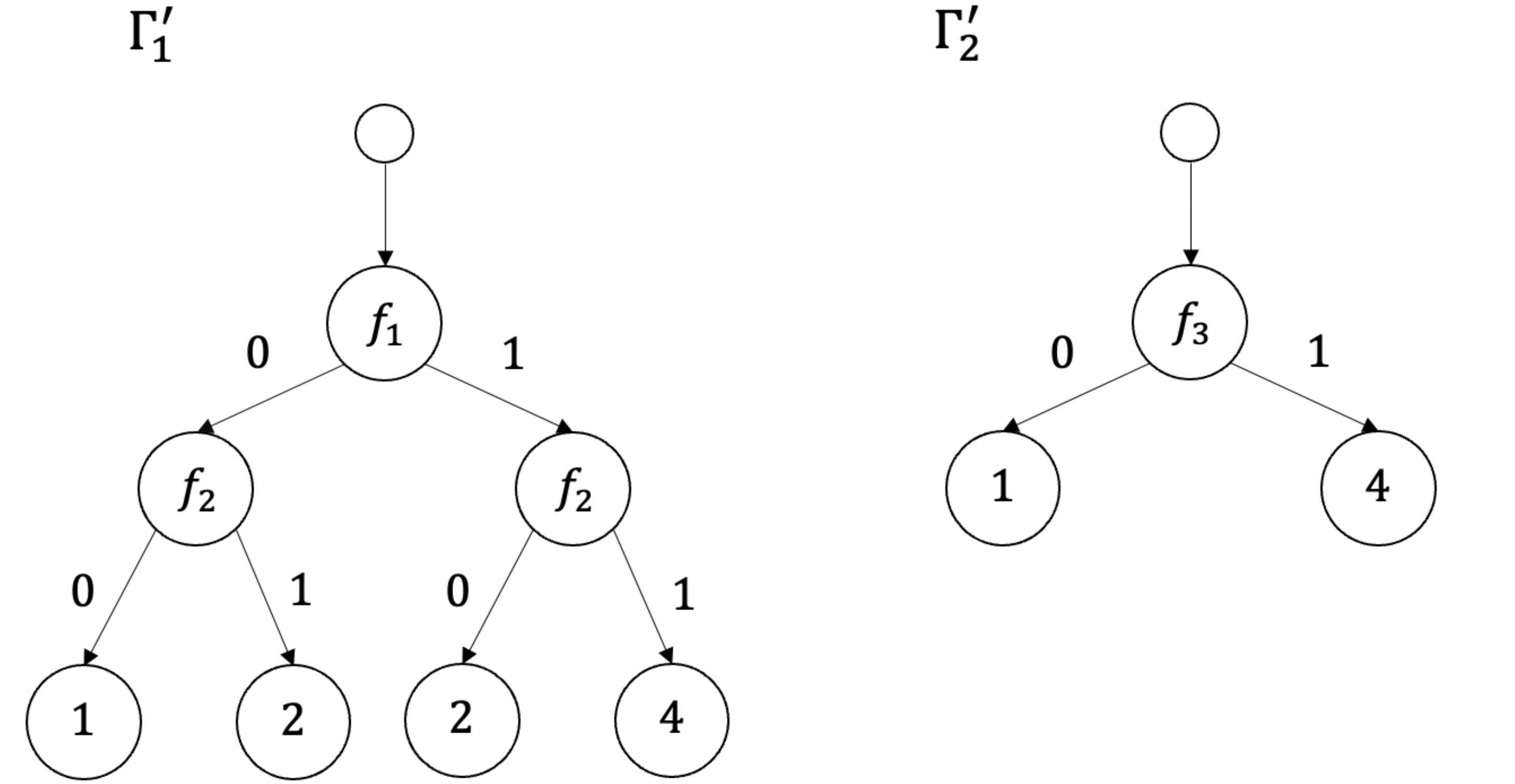}
  \caption{Deterministic decision trees $\Gamma_{1}^{\prime}$ and $\Gamma_{2}^{\prime}$ for decision tables $T_{1}$ and $T_{2}$ depicted in Fig. \ref{fig:fig1}}
  \label{fig:fig7}
\end{figure}

\begin{example}
\label{example7}
Deterministic decision trees $\Gamma_{1}^{\prime}$ and $\Gamma_{2}^{\prime}$ for decision tables $T_{1}$ and $T_{2}$ shown in Fig. \ref{fig:fig1} are depicted in Fig. \ref{fig:fig7}.
\end{example}

\subsection{Complexity Measures}
Denote by $F^{*}$ the set of all finite words over the alphabet $F$ including the empty word $\lambda$.
\begin{definition}
A \textit{complexity measure} over the set of decision tables $\mathcal{M}_{k}(F)$ is any mapping $\psi: F^{*} \rightarrow \mathbb{N}$.
\end{definition}

\begin{definition}
The complexity measure $\psi$ will be called \textit{limited} if it possesses the following properties:

(a) $\psi\left(\alpha_{1} \alpha_{2}\right) \leq \psi\left(\alpha_{1}\right)+\psi\left(\alpha_{2}\right)$ for any $\alpha_{1}, \alpha_{2} \in F^{*}$.

(b) $\psi\left(\alpha_{1} \alpha_{2} \alpha_{3}\right) \geq \psi\left(\alpha_{1} \alpha_{3}\right)$ for any $\alpha_{1}, \alpha_{2}, \alpha_{3} \in F^{*}$.

(c) For any $\alpha \in F^{*}$, the inequality $\psi(\alpha) \geq |\alpha|$ holds, where $|\alpha|$ is the length of $\alpha$.

\end{definition}


We extend an arbitrary complexity measure $\psi$ onto the set $\mathcal{T}_{k}(F)$ in the following way. Let $\Gamma\in \mathcal{T}_{k}(F)$. Then  $\psi(\Gamma)=\max \{\psi(\varphi(\xi)): \xi \in \operatorname{Path}(\Gamma)\}$, where $\varphi(\xi)=\lambda$ if $\pi(\xi)=\lambda$ and $\varphi(\xi)=f_{1}\cdots f_{m}$ if $\pi(\xi)=(f_{1}, \delta_{1})\cdots (f_{m}, \delta_{m})$. The value $\psi(\Gamma)$ will be called the \textit{complexity of the decision tree $\Gamma$}.

We now consider an example of a complexity measure. Let $w: F \rightarrow \mathbb{N} \setminus \{0\}$. We define the function $\psi^{w}: F^{*} \rightarrow \mathbb{N}$ in the following way: $\psi^{w}(\alpha)=0$ if $\alpha=\lambda$ and $\psi^{w}(\alpha)=\sum_{i=1}^{m} w\left(f_{i}\right)$ if $\alpha=f_{1} \cdots f_{m}$. The function $\psi^{w}$ is a limited complexity measure over $\mathcal{M}_{k}(F)$ and it is called a \textit{weighted depth}. If $w \equiv 1$, then the function $\psi^{w}$ is called the \textit{depth} and is denoted by $h$.

Let $\psi$ be a complexity measure over $\mathcal{M}_{k}(F)$ and $T$ be a decision table from $\mathcal{M}_{k}(F)$ in which rows are labeled with attributes $f_{1},\ldots, f_{n}$. The value $\psi^{i}(T)=\psi\left(f_{1} \cdots f_{n}\right)$ will be called the \textit{complexity of the decision table} $T$. We denote by $\psi^{d}(T)$ the minimum complexity of a deterministic decision tree for the table $T$. We denote by $\psi^{a}(T)$ the minimum complexity of a nondeterministic decision tree for the table $T$.

\subsection{Information Systems}
Let $A$ be a nonempty set and $F$ be a nonempty set of functions from $A$ to $E_{k}$.
\begin{definition}
Functions from $F$ are called \textit{attributes} and the pair $U=(A, F)$ is called an \textit{information system}.
\end{definition}

\begin{definition}
A \textit{problem over} $U$ is any $(n+1)$-tuple $z=(\nu, f_{1},\ldots,f_{n})$, where $n\in\mathbb{N}\setminus\{0\}$, $\nu:E_{k}^{n}\rightarrow\mathcal{P}(\mathbb{N})$, and $f_{1},\ldots, f_{n}\in F$.
\end{definition}

The problem $z$ can be interpreted as a problem of searching for at least one number from the set $z(a)=\nu(f_{1}(a),\ldots,f_{n}(a))$ for a given $a\in A$. We denote by $Probl(U)$ the set of problems over the information system $U$.

We correspond to the problem $z$ a decision table $T(z)\in\mathcal{M}_{k}(F)$. This table has $n$ columns labeled with attributes $f_{1},\ldots,f_{n}$. A tuple $\bar{\delta}=(\delta_{1},\ldots,\delta_{n})\in E_{k}^{n}$ is a row of the table $T(z)$ if and only if the system of equations
$$
\{f_{1}(x)=\delta_{1},\ldots,f_{n}(x)=\delta_{n}\}
$$
has a solution from the set $A$. This row is labeled with the set of decisions $\nu(\bar{\delta})$. Let $Tab(U)=\{T(z):z\in Probl(U)\}$. One can show the set $Tab(U)$ is a closed class of decision tables.

Closed classes of decision tables based on information systems are the most natural examples of closed classes. However, the notion of a closed class is essentially wider. In particular, the union $Tab(U_{1})\cup Tab(U_{2})$, where $U_{1}$ and $U_{2}$ are information systems, is a closed class, but generally, we cannot find an information system $U$ such that $Tab(U)=Tab(U_{1})\cup Tab(U_{2})$.

\subsection{Types of T-Pairs}
First, we define the notion of t-pair.
\begin{definition}
A pair $(\mathcal{C}, \psi)$ where $\mathcal{C}$ is a closed class of decision tables from $\mathcal{M}_{k}(F)$ and $\psi$ is a complexity measure over $\mathcal{M}_{k}(F)$ will be called a \textit{test-pair} (or, $t$\textit{-pair}, in short). If $\psi$ is a limited complexity measure then t-pair $(\mathcal{C}, \psi)$ will be called a \textit{limited t-pair}.
\end{definition}

Let $(\mathcal{C}, \psi)$ be a t-pair. We have three parameters $\psi^{i}(T), \psi^{d}(T)$ and $\psi^{a}(T)$ for any decision table $T \in \mathcal{C}$. We now define functions that describe relationships among these parameters. Let $b, c \in\{i, d, a\}$.
\begin{definition}
We define partial functions $\mathcal{U}_{\mathcal{C} \psi}^{b c}: \mathbb{N} \rightarrow \mathbb{N}$ and $\mathcal{L}_{\mathcal{C} \psi}^{b c}: \mathbb{N} \rightarrow \mathbb{N}$ by
$$
\begin{aligned}
&\mathcal{U}_{\mathcal{C} \psi}^{b c}(n)=\max \left\{\psi^{b}(T): T \in \mathcal{C}, \psi^{c}(T) \leq n\right\}, \\
&\mathcal{L}_{\mathcal{C} \psi}^{b c}(n)=\min \left\{\psi^{b}(T): T \in \mathcal{C}, \psi^{c}(T) \geq n\right\}.
\end{aligned}
$$
\end{definition}
If the value $\mathcal{U}_{\mathcal{C} \psi}^{b c}(n)$ is definite, then it is the unimprovable upper bound on the values $\psi^{b}(T)$ for tables $T \in \mathcal{C}$ satisfying $\psi^{c}(T) \leq n$. If the value $\mathcal{L}_{\mathcal{C} \psi}^{b c}(n)$ is definite then it is the unimprovable lower bound on the values $\psi^{b}(T)$ for tables $T \in \mathcal{C}$ satisfying $\psi^{c}(T) \geq n$.

Let $g$ be a partial function from $\mathbb{N}$ to $\mathbb{N}$. We denote by $\operatorname{Dom}(g)$ the domain of $g$. Denote $\operatorname{Dom}^{+}(g)=\{n: n \in \operatorname{Dom}(g), g(n) \geq n\}$ and $\operatorname{Dom}^{-}(g)=\{n: n \in$ $\operatorname{Dom}(g), g(n) \leq n\}$.
\begin{definition}
Now we define the value $\operatorname{typ}(g) \in\{\alpha, \beta, \gamma, \delta, \epsilon\}$ called the \textit{type} of $g$.

\begin{itemize}
  \item If $\operatorname{Dom}(g)$ is an infinite set and $g$ is a bounded from above function, then $\operatorname{typ}(g)=\alpha$.

  \item If $\operatorname{Dom}(g)$ is an infinite set, $\operatorname{Dom}^{+}(g)$ is a finite set, and $g$ is an unbounded from above function, then $\operatorname{typ}(g)=\beta$.

  \item If both sets $\operatorname{Dom}^{+}(g)$ and $\operatorname{Dom}^{-}(g)$ are infinite, then $\operatorname{typ}(g)=\gamma$.

  \item If $\operatorname{Dom}(g)$ is an infinite set and $\operatorname{Dom}^{-}(g)$ is a finite set, then $\operatorname{typ}(g)=\delta$.

  \item If $\operatorname{Dom}(g)$ is a finite set, then $\operatorname{typ}(g)=\epsilon$.

\end{itemize}
\end{definition}

\begin{example}
    One can show that $\operatorname{typ}(1)=\alpha$, $\operatorname{typ}(\lceil \log_{2} n \rceil)=\beta$, $\operatorname{typ}(n)=\gamma$, $\operatorname{typ}(n^{2})=\delta$, and $\operatorname{typ}(\frac{1}{\lfloor 1/n \rfloor})=\epsilon$.
\end{example}
\begin{definition}
We now define the table $\operatorname{typ}(\mathcal{C}, \psi)$, which will be called the \textit{type of t-pair} $(\mathcal{C}, \psi)$. This is a table with three rows and three columns in which rows from the top to the bottom and columns from the left to the right are labeled with indices $i, d, a$. The pair $\operatorname{typ}(\mathcal{L}_{\mathcal{C} \psi}^{b c}) \operatorname{typ}(\mathcal{U}_{\mathcal{C} \psi}^{b c})$ is in the intersection of the row with index $b \in\{i, d, a\}$ and the column with index $c \in\{i, d, a\}$.
\end{definition}

\section{Main Results}  \label{S3}
The main problem investigated in this paper is finding all types of t-pairs and limited t-pairs. The solution to this problem describes all possible (in terms of functions $\mathcal{U}_{\mathcal{C} \psi}^{b c}, \mathcal{L}_{\mathcal{C} \psi}^{b c}$ types, $b, c \in\{i, d, a\}$) relationships among the complexity of decision tables, the minimum complexity of nondeterministic decision trees for them, and the minimum complexity of deterministic decision trees for these tables. We now define seven tables:
\\
\\
\begin{minipage}[c]{0.23\textwidth}
\begin{center}
$T_{1}=$
\begin{tabular}{ |c|ccc| }
 \hline
  & $i$ & $d$ & $a$\\
  \hline
 $i$ & $\epsilon\alpha$ & $\epsilon\alpha$  & $\epsilon\alpha$ \\
 $d$ & $\epsilon\alpha$  & $\epsilon\alpha$  &$\epsilon\alpha$ \\
 $a$ & $\epsilon\alpha$  & $\epsilon\alpha$  &$\epsilon\alpha$ \\
 \hline
\end{tabular}
\end{center}
\end{minipage}
\begin{minipage}[c]{0.23\textwidth}
\begin{center}
$T_{2}=$
\begin{tabular}{ |c|ccc| }
 \hline
  & $i$ & $d$ & $a$\\
  \hline
 $i$ & $\gamma\gamma$ & $\epsilon\epsilon$  & $\epsilon\epsilon$ \\
 $d$ & $\alpha\alpha$  & $\epsilon\alpha$  &$\epsilon\alpha$ \\
 $a$ & $\alpha\alpha$  & $\epsilon\alpha$  &$\epsilon\alpha$ \\
 \hline
\end{tabular}
\end{center}
\end{minipage}
\begin{minipage}[c]{0.23\textwidth}
\begin{center}
$T_{3}=$
\begin{tabular}{ |c|ccc| }
 \hline
  & $i$ & $d$ & $a$\\
  \hline
 $i$ & $\gamma\gamma$ & $\delta\epsilon$  & $\epsilon\epsilon$ \\
 $d$ & $\alpha\beta$  & $\gamma\gamma$  &$\epsilon\epsilon$ \\
 $a$ & $\alpha\alpha$  & $\alpha\alpha$  &$\epsilon\alpha$ \\
 \hline
\end{tabular}
\end{center}
\end{minipage}
\begin{minipage}[c]{0.23\textwidth}
\begin{center}
$T_{4}=$
\begin{tabular}{ |c|ccc| }
 \hline
  & $i$ & $d$ & $a$\\
  \hline
 $i$ & $\gamma\gamma$ & $\gamma\epsilon$  & $\epsilon\epsilon$ \\
 $d$ & $\alpha\gamma$  & $\gamma\gamma$  &$\epsilon\epsilon$ \\
 $a$ & $\alpha\alpha$  & $\alpha\alpha$  &$\epsilon\alpha$ \\
 \hline
\end{tabular}
\end{center}
\end{minipage}
\\
\\

\begin{minipage}[c]{0.3\textwidth}
\centering
$T_{5}=$
\begin{tabular}{ |c|ccc| }
 \hline
  & $i$ & $d$ & $a$\\
  \hline
 $i$ & $\gamma\gamma$ & $\gamma\epsilon$  & $\gamma\epsilon$ \\
 $d$ & $\alpha\gamma$  & $\gamma\gamma$  &$\gamma\gamma$ \\
 $a$ & $\alpha\gamma$  & $\gamma\gamma$  &$\gamma\gamma$ \\
 \hline
\end{tabular}
\end{minipage}
\begin{minipage}[c]{0.3\textwidth}
\centering
$T_{6}=$
\begin{tabular}{ |c|ccc| }
 \hline
  & $i$ & $d$ & $a$\\
  \hline
 $i$ & $\gamma\gamma$ & $\gamma\epsilon$  & $\gamma\epsilon$ \\
 $d$ & $\alpha\gamma$  & $\gamma\gamma$  &$\gamma\delta$ \\
 $a$ & $\alpha\gamma$  & $\beta\gamma$  &$\gamma\gamma$ \\
 \hline
\end{tabular}
\end{minipage}
\begin{minipage}[c]{0.3\textwidth}
\centering
$T_{7}=$
\begin{tabular}{ |c|ccc| }
 \hline
  & $i$ & $d$ & $a$\\
  \hline
 $i$ & $\gamma\gamma$ & $\gamma\epsilon$  & $\gamma\epsilon$ \\
 $d$ & $\alpha\gamma$  & $\gamma\gamma$  &$\gamma\epsilon$ \\
 $a$ & $\alpha\gamma$  & $\alpha\gamma$  &$\gamma\gamma$ \\
 \hline
\end{tabular}
\end{minipage}
\\
\\

\begin{theorem}
\label{theorem1}
For any $t$-pair $(\mathcal{C}, \psi)$, the relation $\operatorname{typ}(\mathcal{C}, \psi) \in\left\{T_{1}, T_{2}, T_{3}, T_{4}, T_{5}\right.$, $\left.T_{6}, T_{7}\right\}$ holds. For any $i \in\{1,2,3,4,5,6,7\}$, there exists a t-pair $(\mathcal{C}, \psi)$ such that $\operatorname{typ}(\mathcal{C}, \psi)=T_{i}$.
\end{theorem}

\begin{theorem}
\label{theorem2}
For any limited t-pair $(\mathcal{C}, \psi)$, the relation $\operatorname{typ}(\mathcal{C}, \psi) \in\left\{T_{2}, T_{3}, T_{5}\right.$, $\left.T_{6}, T_{7}\right\}$ holds. For any $i \in\{2,3,5,6,7\}$, there exists a limited t-pair $(\mathcal{C}, h)$ such that $\operatorname{typ}(\mathcal{C}, h)=T_{i}$.
\end{theorem}



\section{Possible Upper Types of T-Pairs}  \label{S4}
We begin our study by considering the upper type of t-pair, which is a simpler object than the type of t-pair.
\begin{definition}
Let $(\mathcal{C}, \psi)$ be a t-pair. We now define table $\operatorname{typ}_{u}(\mathcal{C}, \psi)$, which will be called the upper type of t-pair $(\mathcal{C}, \psi)$. This is a table with three rows and three columns in which rows from the top to the bottom and columns from the left to the right are labeled with indices $i, d, a$. The value $\operatorname{typ}(\mathcal{U}_{\mathcal{C} \psi}^{b c})$ is in the intersection of the row with index $b \in\{i, d, a\}$ and the column with index $c \in\{i, d, a\}$. The table $\operatorname{typ}_{u}(\mathcal{C}, \psi)$ will be called the \textit{upper type of t-pair} $(\mathcal{C}, \psi)$.
\end{definition}
In this section, all possible upper types of t-pairs are enumerated. We now define seven tables:\\
\\
\begin{minipage}[c]{0.23\textwidth}
\begin{center}
$t_{1}=$
\begin{tabular}{ |c|ccc| }
 \hline
  & $i$ & $d$ & $a$\\
  \hline
 $i$ & $\alpha$ & $\alpha$  & $\alpha$ \\
 $d$ & $\alpha$  & $\alpha$  &$\alpha$ \\
 $a$ & $\alpha$  & $\alpha$  &$\alpha$ \\
 \hline
\end{tabular}
\end{center}
\end{minipage}
\begin{minipage}[c]{0.23\textwidth}
\begin{center}
$t_{2}=$
\begin{tabular}{ |c|ccc| }
 \hline
  & $i$ & $d$ & $a$\\
  \hline
 $i$ & $\gamma$ & $\epsilon$  & $\epsilon$ \\
 $d$ & $\alpha$  & $\alpha$  &$\alpha$ \\
 $a$ & $\alpha$  & $\alpha$  &$\alpha$ \\
 \hline
\end{tabular}
\end{center}
\end{minipage}
\begin{minipage}[c]{0.23\textwidth}
\begin{center}
$t_{3}=$
\begin{tabular}{ |c|ccc| }
 \hline
  & $i$ & $d$ & $a$\\
  \hline
 $i$ & $\gamma$ & $\epsilon$  & $\epsilon$ \\
 $d$ & $\beta$  & $\gamma$  &$\epsilon$ \\
 $a$ & $\alpha$  & $\alpha$  &$\alpha$ \\
 \hline
\end{tabular}
\end{center}
\end{minipage}
\begin{minipage}[c]{0.23\textwidth}
\begin{center}
$t_{4}=$
\begin{tabular}{ |c|ccc| }
 \hline
  & $i$ & $d$ & $a$\\
  \hline
 $i$ & $\gamma$ & $\epsilon$  & $\epsilon$ \\
 $d$ & $\gamma$  & $\gamma$  &$\epsilon$ \\
 $a$ & $\alpha$  & $\alpha$  &$\alpha$ \\
 \hline
\end{tabular}
\end{center}
\end{minipage}
\\
\\

\begin{minipage}[c]{0.3\textwidth}
\centering
$t_{5}=$
\begin{tabular}{ |c|ccc| }
 \hline
  & $i$ & $d$ & $a$\\
  \hline
 $i$ & $\gamma$ & $\epsilon$  & $\epsilon$ \\
 $d$ & $\gamma$  & $\gamma$  &$\gamma$ \\
 $a$ & $\gamma$  & $\gamma$  &$\gamma$ \\
 \hline
\end{tabular}
\end{minipage}
\begin{minipage}[c]{0.3\textwidth}
\centering
$t_{6}=$
\begin{tabular}{ |c|ccc| }
 \hline
  & $i$ & $d$ & $a$\\
  \hline
 $i$ & $\gamma$ & $\epsilon$  & $\epsilon$ \\
 $d$ & $\gamma$  & $\gamma$  &$\delta$ \\
 $a$ & $\gamma$  & $\gamma$  &$\gamma$ \\
 \hline
\end{tabular}
\end{minipage}
\begin{minipage}[c]{0.3\textwidth}
\centering
$t_{7}=$
\begin{tabular}{ |c|ccc| }
 \hline
  & $i$ & $d$ & $a$\\
  \hline
 $i$ & $\gamma$ & $\epsilon$  & $\epsilon$ \\
 $d$ & $\gamma$  & $\gamma$  &$\epsilon$ \\
 $a$ & $\gamma$  & $\gamma$  &$\gamma$ \\
 \hline
\end{tabular}
\end{minipage}
\\
\\

\begin{proposition}
\label{proposition1}
For any $t$-pair $(\mathcal{C}, \psi)$, the relation $\operatorname{typ}_{u}(\mathcal{C}, \psi) \in\left\{t_{1}, t_{2}, t_{3}, t_{4}\right.$, $\left.t_{5}, t_{6}, t_{7}\right\}$ holds.
\end{proposition}

\begin{proposition}
\label{proposition2}
For any limited t-pair $(\mathcal{C}, \psi)$, the relation $\operatorname{typ}_{u}(\mathcal{C}, \psi) \in\left\{t_{2}, t_{3}\right.$, $\left.t_{5}, t_{6}, t_{7}\right\}$ holds.
\end{proposition}

We divide the proofs of the propositions into a sequence of lemmas. 

\begin{lemma}
\label{lemma1}
Let $T$ be a decision table from a set of decision tables $\mathcal{M}_{k}(F)$ and $\psi$ be a complexity measure over $\mathcal{M}_{k}(F)$. Then the inequalities $\psi^{a}(T) \leq \psi^{d}(T) \leq \psi^{i}(T)$ hold.
\end{lemma}

\begin{proof}
Let columns of the table $T$ be labeled with attributes $f_{1},\ldots,f_{n}$. It is not difficult to construct a deterministic decision tree $\Gamma_{0}$ for the table $T$, which sequentially computes values of attributes $f_{1}, \ldots, f_{n}$. Evidently, $\psi(\Gamma_{0})=\psi^{i}(T)$. Therefore $\psi^{d}(T) \leq \psi^{i}(T)$. If a decision tree $\Gamma$ is a deterministic decision tree for $T$, then $\Gamma$ is a nondeterministic decision tree for $T$. Therefore $\psi^{a}(T) \leq \psi^{d}(T)$.
\qed \end{proof}

Let $(\mathcal{C}, \psi)$ be a t-pair, $n \in \mathbb{N}$ and $b, c \in\{i, d, a\}$. The notation $\mathcal{U}_{\mathcal{C} \psi}^{b c}(n)=\infty$ means that the set $X=\{\psi^{b}(T): T \in \mathcal{C}, \psi^{c}(T) \leq n\}$ is infinite. The notation $\mathcal{U}_{\mathcal{C} \psi}^{b c}(n)=\varnothing$ means that the set $X$ is empty. Evidently, if $\mathcal{U}_{\mathcal{C} \psi}^{b c}(n)=\infty$, then $\mathcal{U}_{\mathcal{C} \psi}^{b c}(n+1)=\infty$. It is not difficult to prove the following statement.

\begin{lemma}
\label{lemma2}
Let $(\mathcal{C}, \psi)$ be a t-pair and $b, c \in\{i, d, a\}$. Then

(a) If there exists $n \in \mathbb{N}$ such that $\mathcal{U}_{\mathcal{C} \psi}^{b c}(n)=\infty$, then $\operatorname{typ}(\mathcal{U}_{\mathcal{C} \psi}^{b c})=\epsilon$.

(b) If there is no $n \in \mathbb{N}$ such that $\mathcal{U}_{\mathcal{C} \psi}^{b c}(n)=\infty$, then $\operatorname{Dom}(\mathcal{U}_{\mathcal{C} \psi}^{b c})=\{n: n \in$ $\left.\mathbb{N}, n \geq n_{0}\right\}$, where $n_{0}=\min \{\psi^{c}(T): T \in \mathcal{C}\}$.
\end{lemma}

Let $(\mathcal{C}, \psi)$ be a t-pair and $b, c, e, f \in\{i, d, a\}$. The notation $\mathcal{U}_{\mathcal{C} \psi}^{b c} \triangleleft \mathcal{U}_{\mathcal{C} \psi}^{e f}$ means that, for any $n \in \mathbb{N}$, the following statements hold:

(a) If the value $\mathcal{U}_{\mathcal{C} \psi}^{b c}(n)$ is definite then either $\mathcal{U}_{\mathcal{C} \psi}^{e f}(n)=\infty$ or the value $\mathcal{U}_{\mathcal{C} \psi}^{e f}(n)$ is definite and the inequality $\mathcal{U}_{\mathcal{C} \psi}^{b c}(n) \leq \mathcal{U}_{\mathcal{C} \psi}^{e f}(n)$ holds.

(b) If $\mathcal{U}_{\mathcal{C} \psi}^{b c}(n)=\infty$, then $\mathcal{U}_{\mathcal{C} \psi}^{e f}(n)=\infty$.

Let $\preceq$ be a linear order on the set $\{\alpha, \beta, \gamma, \delta, \epsilon\}$ such that $\alpha \preceq \beta \preceq \gamma \preceq \delta \preceq \epsilon$.

\begin{lemma}
\label{lemma3}
Let $(\mathcal{C}, \psi)$ be a t-pair. Then $\operatorname{typ}(\mathcal{U}_{\mathcal{C} \psi}^{b i}) \preceq \operatorname{typ}(\mathcal{U}_{\mathcal{C} \psi}^{b d}) \preceq \operatorname{typ}(\mathcal{U}_{\mathcal{C} \psi}^{b a})$ and  $\operatorname{typ}(\mathcal{U}_{\mathcal{C} \psi}^{a b}) \preceq \operatorname{typ}(\mathcal{U}_{\mathcal{C} \psi}^{d b}) \preceq \operatorname{typ}(\mathcal{U}_{\mathcal{C} \psi}^{i b})$ for any $b \in\{i, d, a\}$.
\end{lemma}

\begin{proof}
From the definition of the functions $\mathcal{U}_{\mathcal{C} \psi}^{b c}, b, c \in\{i, d, a\}$, and from Lemma \ref{lemma1} it follows that $\mathcal{U}_{\mathcal{C} \psi}^{b i} \triangleleft \mathcal{U}_{\mathcal{C} \psi}^{b d} \triangleleft \mathcal{U}_{\mathcal{C} \psi}^{b a}$ and $\mathcal{U}_{\mathcal{C} \psi}^{a b} \triangleleft \mathcal{U}_{\mathcal{C} \psi}^{d b} \triangleleft \mathcal{U}_{\mathcal{C} \psi}^{i b}$ for any $b \in\{i, d, a\}$. Using these relations and Lemma \ref{lemma2} we obtain the statement of the lemma.
\qed \end{proof}

\begin{lemma}
\label{lemma4}
Let $(\mathcal{C}, \psi)$ be a t-pair and $b, c \in\{i, d, a\}$. Then

(a) $\operatorname{typ}(\mathcal{U}_{\mathcal{C} \psi}^{b c})=\alpha$ if and only if the function $\psi^{b}$ is bounded from above on the closed class $\mathcal{C}$.

(b) If the function $\psi^{b}$ is unbounded from above on $\mathcal{C}$, then $\operatorname{typ}(\mathcal{U}_{\mathcal{C} \psi}^{b b})=\gamma$.
\end{lemma}

\begin{proof}
The statement (a) is obvious. (b) Let the function $\psi^{b}$ be unbounded from above on $\mathcal{C}$. One can show that in this case the equality $\mathcal{U}_{\mathcal{C} \psi}^{b b}(n)=n$ holds for infinitely many $n \in \mathbb{N}$. Therefore $\operatorname{typ}(\mathcal{U}_{\mathcal{C} \psi}^{b b})=\gamma$.
\qed \end{proof}

\begin{corollary}
\label{corollary1}
Let $(\mathcal{C}, \psi)$ be a t-pair and $b \in\{i, d, a\}$. Then $\operatorname{typ}(\mathcal{U}_{\mathcal{C} \psi}^{b b}) \in\{\alpha, \gamma\}$.
\end{corollary}

\begin{lemma}
\label{lemma5}
Let $(\mathcal{C}, \psi)$ be a t-pair and $\operatorname{typ}(\mathcal{U}_{\mathcal{C} \psi}^{i i}) \neq \alpha$. Then $$\operatorname{typ}(\mathcal{U}_{\mathcal{C} \psi}^{i d})=\operatorname{typ}(\mathcal{U}_{\mathcal{C} \psi}^{i a})=\epsilon.$$
\end{lemma}

\begin{proof}
Using Lemma \ref{lemma4}, we conclude that the function $\psi^{i}$ is unbounded from above on $\mathcal{C}$. Let $m \in \mathbb{N}$. Then there exists a decision table $T \in$ $\mathcal{C}$ for which the inequality $\psi^{i}(T) \geq m$ holds. Let us consider a degenerate decision table $T^{\prime}\in \mathcal{C}$ obtained from $T$ by replacing the sets of decisions attached to rows by the set $\{0\}$. It is clear that $\psi^{i}(T^{\prime}) \geq m$. Let $\Gamma$ be a decision tree, which consists of the root, the terminal node labeled with $0$, and the edge connecting these two nodes. One can show that $\Gamma$ is a deterministic decision tree for the table $T^{\prime}$. Therefore $\psi^{a}(T^{\prime}) \leq \psi^{d}(T^{\prime}) \leq \psi(\Gamma)=\psi(\lambda)$. Taking into account that $m$ is an arbitrary number from $\mathbb{N}$, we obtain $\mathcal{U}_{\mathcal{C} \psi}^{i d}(\psi(\lambda))=\infty$ and $\mathcal{U}_{\mathcal{C} \psi}^{i a}(\psi(\lambda))=\infty$. Using Lemma \ref{lemma2}, we conclude that $\operatorname{typ}(\mathcal{U}_{\mathcal{C} \psi}^{i d})=\operatorname{typ}(\mathcal{U}_{\mathcal{C} \psi}^{i a})=\epsilon$.
\qed \end{proof}
\begin{example}
\label{example9}
Let us consider a t-pair $(\mathcal{C}_{0}, h)$, where $\mathcal{C}_{0}$ is closed class described in Example \ref{example5}. It is clear that the function $h^{i}$ is unbounded from above on $\mathcal{C}_{0}$ and the functions $h^{a}$ and $h^{d}$ are bounded from above on $\mathcal{C}_{0}$. Using Lemma \ref{lemma4}, we obtain that $\operatorname{typ}(\mathcal{U}_{\mathcal{C}_{0} h}^{ab})=\operatorname{typ}(\mathcal{U}_{\mathcal{C}_{0} h}^{db})=\alpha$ for any $b\in\{i, d, a\}$ and $\operatorname{typ}(\mathcal{U}_{\mathcal{C}_{0} h}^{ii})=\gamma$. By Lemma \ref{lemma5}, $\operatorname{typ}(\mathcal{U}_{\mathcal{C}_{0} h}^{id})=\operatorname{typ}(\mathcal{U}_{\mathcal{C}_{0} h}^{ia})=\epsilon$. Therefore, $\operatorname{typ}_{u}(\mathcal{C}_{0}, h)=t_{2}$.
\end{example}
\begin{lemma}
\label{lemma6}
Let $(\mathcal{C}, \psi)$ be a t-pair. Then $\operatorname{typ}(\mathcal{U}_{\mathcal{C} \psi}^{a i}) \in\{\alpha, \gamma\}$.
\end{lemma}

\begin{proof}
Using Lemma \ref{lemma3} and Corollary \ref{corollary1}, we obtain $\operatorname{typ}(\mathcal{U}_{\mathcal{C} \psi}^{a i}) \in\{\alpha, \beta, \gamma\}$. By Lemma \ref{lemma2}, $\operatorname{Dom}(\mathcal{U}_{\mathcal{C} \psi}^{a i})=\{n:n\in \mathbb{N}, n\geq n_{0}\}$ for some $n_{0}\in \mathbb{N}$. Set $D=\operatorname{Dom}(\mathcal{U}_{\mathcal{C} \psi}^{a i})$. Assume that $\operatorname{typ}(\mathcal{U}_{\mathcal{C} \psi}^{a i})=\beta$. Then there exists $m \in D$ such that $\mathcal{U}_{\mathcal{C} \psi}^{a i}(n)<n$ for any $n \in D, n>m$. Let us prove by induction on $n$ that, for any decision table $T$ from $\mathcal{C}$, if $\psi^{i}(T) \leq n$, then $\psi^{a}(T) \leq m_{0}$, where $m_{0}=\max \{m, \psi(\lambda)\}$. Using Lemma \ref{lemma1}, we conclude that under the condition $n \leq m$ the considered statement holds. Let it hold for some $n, n \geq m$. Let us show that this statement holds for $n+1$ too. Let $T \in \mathcal{C}$, $\psi^{i}(T) \leq n+1$ and columns of the table $T$ be labeled with attributes $f_{i_{1}},\ldots,f_{i_{k}}$. Since $n+1>m$, we obtain $\psi^{a}(T) \leq n$. Let $\Gamma$ be a nondeterministic decision tree for the table $T$ and  $\psi(\Gamma)=\psi^{a}(T)$. Assume that in $\Gamma$ there exists a complete path $\xi$ in which there are no worker nodes. In this case, a decision tree, which consists of the root, the terminal node labeled with $\tau(\xi)$ and the edge connecting these two nodes is a nondeterministic decision tree for the table $T$. Therefore $\psi^{a}(T) \leq \psi(\lambda) \leq m_{0}$. Assume now that each complete path in the decision tree $\Gamma$ contains a worker node. Let $\xi \in \operatorname{Path}(\Gamma), \Delta(T\pi(\xi))\neq \varnothing$, $\xi=v_{0}, d_{0}, \ldots, v_{p}, d_{p}, v_{p+1}$ and, for $i=1, \ldots, p$, the node $v_{i}$ be labeled with the attribute $f_{i}$, and the edge $d_{i}$ be labeled with the number $\delta_{i}$. 
Let the decision table $T^{\prime}$ be obtained from the decision table $T$ by operations of permutation of columns and duplication of columns so that its columns are labeled with attributes $f_{1},\ldots,f_{p},f_{i_{1}},\ldots,f_{i_{k}}$. We obtain the decision table $T^{\prime\prime}$ from $T^{\prime}$ by removal the last $k$ columns. Let us denote by $T_{\xi}$ the decision table obtained from $T^{\prime\prime}$ by changing the set of decisions corresponding to the row $(\delta_{1},\ldots,\delta_{p})$ with $\{\tau(\xi)\}$, and for the remaining rows with $\{\tau(\xi)+1\}$.
It is clear that $\psi^{i}\left(T_{\xi}\right) \leq n$. Using the inductive hypothesis, we conclude that there exists a nondeterministic decision tree $\Gamma_{\xi}$ for the table $T_{\xi}$ such that $\psi(\Gamma_{\xi}) \leq m_{0}$. We denote by $\tilde{\Gamma}_{\xi}$ a tree obtained from $\Gamma_{\xi}$ by removal of all nodes and edges that satisfy the following condition: there is no a complete path $\xi^{\prime}$ in $\Gamma_{\xi}$, which contains this node or edge and for which $\tau\left(\xi^{\prime}\right)=\tau(\xi)$. Let $\{\xi: \xi \in \operatorname{Path}(\Gamma), \Delta(T\pi(\xi))\neq \varnothing\}=\left\{\xi_{1}, \ldots, \xi_{r}\right\}$. Let us identify the roots of the trees $\tilde{\Gamma}_{\xi_{1}}, \ldots, \tilde{\Gamma}_{\xi_{r}}$. We denote by $G$ the obtained tree. It is not difficult to show that $G$ is a nondeterministic decision tree for the table $T$ and $\psi(G) \leq m_{0}$. Thus, the considered statement holds. Using Lemma \ref{lemma4}, we conclude that $\operatorname{typ}(\mathcal{U}_{\mathcal{C} \psi}^{a i})=\alpha$. The obtained contradiction shows that $\operatorname{typ}(\mathcal{U}_{\mathcal{C} \psi}^{a i}) \in\{\alpha, \gamma\}$.
\qed \end{proof}

Let $T$ be a decision table from $\mathcal{M}_{k}(F)$. We now give definitions of parameters $N(T)$ and $M(T)$ of the table $T$. 
\begin{definition}
    We denote by $N(T)$ the number of rows in the table $T$.
\end{definition}
\begin{definition}
    Let columns of table $T$ be labeled with attributes $f_{1},\ldots,f_{n}\in F$. We now define the parameter $M(T)$. If table $T$ is degenerate, then $M(T)=0$. Let now $T$ be a nondegenerate table and $\bar{\delta}=(\delta_{1},\ldots,\delta_{n})\in E_{k}^{n}$. Then $M(T, \bar{\delta})$ is the minimum natural $m$ such that there exist attributes $f_{i_1},\ldots, f_{i_m}\in At(T)$ for which $T(f_{i_1}, \delta_{i_1})\cdots(f_{i_m}, \delta_{i_m})$ is a degenerate table. We denote $M(T)=\max\{M(T, \bar{\delta}):\bar{\delta}\in E_{k}^{n}\}$.
\end{definition}

The following statement follows immediately from Theorem 3.5 \cite{Moshkov05}.
\begin{lemma}
    Let $T$ be a nonempty decision table from $\mathcal{M}_{k}(F)$ in which each row is labeled with a set containing only one decision. Then
    $$
    h^{d}(T)\leq M(T)\log_{2} N(T).
    $$
    \label{lemma_h}
\end{lemma}
\begin{lemma}
\label{lemma7}
Let $(\mathcal{C}, \psi)$ be a limited t-pair and $\operatorname{typ}(\mathcal{U}_{\mathcal{C} \psi}^{a i})=\alpha$. Then $\operatorname{typ}(\mathcal{U}_{\mathcal{C} \psi}^{d i}) \in$ $\{\alpha, \beta\}$.
\end{lemma}

\begin{proof}
Using Lemma \ref{lemma4}, we conclude that there exists $r \in \mathbb{N}$ such that the inequality $\psi^{a}(T) \leq r$ holds for any table $T \in \mathcal{C}$.

Let $T$ be a nonempty table from $\mathcal{C}$ in which columns are labeled with attributes $f_{1},\ldots,f_{n}$ and $\bar{\delta}=(\delta_{1},\ldots,\delta_{n})\in E_{k}^{n}$. We now show that there exist attributes $f_{i_1},\ldots,f_{i_m}\in At(T)$ such that the subtable $T(\bar{\delta})=T(f_{1},\delta_{1})\cdots(f_{n},\delta_{n})$ is equal to the subtable $T(f_{i_1}, \delta_{i_1})\cdots(f_{i_m}, \delta_{i_m})$ and $m\leq r$ if $\bar{\delta}$ is a row of $T$, and $m\leq r+1$ if $\bar{\delta}$ is not a row of $T$.

Let $\bar{\delta}$ be a row of $T$.
Let us change the set of decisions attached to the row $\bar{\delta}$ with the set $\{1\}$ and for the remaining rows of $T$ with the set $\{0\}$. We denote the obtained table by $T^{\prime}$. It is clear that $T^{\prime}\in \mathcal{C}$. 
Taking into account that $\psi^{a}(T^{\prime}) \leq r$ and the complexity measure $\psi$ has the property (c), it is not difficult to show that there exist attributes $f_{i_1},\ldots,f_{i_m}\in At(T^{\prime})=At(T)$ such that $m\leq r$ and $T^{\prime}(f_{i_1}, \delta_{i_1})\cdots(f_{i_m}, \delta_{i_m})$ contains only the row $\bar{\delta}$. From here it follows that $T(\bar{\delta})=T(f_{i_1}, \delta_{i_1})\cdots(f_{i_m}, \delta_{i_m})$.

Let $\bar{\delta}$ be not a row of $T$. Let us show that there exist attributes $f_{i_1},\ldots, f_{i_m}\in At(T)$ such that $m\leq r+1$ and the subtable $T(f_{i_1}, \delta_{i_1})\cdots(f_{i_m}, \delta_{i_m})$ is empty. If $T(f_1,\delta_1)$ is empty, then the considered statement holds. Otherwise, there exists $q \in\{1, \ldots, n-1\}$ such that the subtable $T(f_{1}, \delta_{1})\cdots(f_{q}, \delta_{q})$ is nonempty but the subtable $T(f_{1}, \delta_{1})\cdots(f_{{q+1}}, \delta_{{q+1}})$ is empty.
We denote by $T^{\prime}$ the table obtained from $T$ by removal of attributes $f_{q+1},\ldots,f_{n}$. It is clear that $T^{\prime}\in \mathcal{C}$ and $(\delta_{1},\ldots,\delta_{q})$ is a row of $T^{\prime}$. According to proven above, there exist attributes $f_{i_1},\ldots,f_{i_p}\in \{f_{1},\ldots,f_{q}\}$ such that $$T^{\prime}(f_{i_1},\delta_{i_1})\cdots(f_{i_p},\delta_{i_p})=T^{\prime}(f_{1}, \delta_{1})\cdots(f_{q}, \delta_{q})$$ and $p\leq r$. Using this fact one can show that $T(f_{i_1},\delta_{i_1})\cdots(f_{i_p},\delta_{i_p})(f_{q+1}, \delta_{q+1})$ is empty and is equal to $T(\bar{\delta})$.

Let $T_{1}\in\mathcal{C}$. We denote by $T_2$ the decision table obtained from $T_1$ by removal of all columns in which all numbers are equal. 
Let columns of $T_{2}$ be labeled with attributes $f_{1}, \ldots, f_{n}$. We now consider the decision table $T_{3}$, which is obtained from $T_{2}$ by changing decisions so that the decision set attached to each row of table $T_{3}$ contains only one decision and, for any two non-equal rows, corresponding decisions are different. It is clear that $T_{3}\in \mathcal{C}$. It is not difficult to show that $\psi^{d}(T_{1}) \leq \psi^{d}(T_{2}) \leq \psi^{d}(T_{3})$.

We now show that the inequality $\psi(f)\leq r$ holds for any attribute $f\in At(T_{3})$. Let us denote by $T^{\prime}$ the decision table obtained from $T_{3}$ by removal of all columns except the column labeled with the attribute $f$. If there is more than one column in $T_{3}$, which is labeled with the attribute $f$, then we keep only one of them. Let the decision table $T_{f}$ be obtained from $T^{\prime}$ by changing the set of decisions for each row $(\delta)$ with the set of decisions $\{\delta\}$. It is clear that $T_{f}\in \mathcal{C}$. Let $\Gamma$ be a nondeterministic decision tree for the table $T_{f}$ and $\psi(\Gamma)=\psi^{a}(T_{f})\leq r$. Since the column $f$ contains different numbers, we have $f \in At(\Gamma)$. Using the property (b) of the complexity measure $\psi$, we obtain $\psi(\Gamma) \geq \psi(f)$. Consequently, $\psi(f) \leq r$. 

Taking into account that, for any $\bar{\delta}\in \Delta(T_{3})$, there exist attributes $f_{i_1},\ldots,$ $f_{i_m}\in \{f_{1},\ldots,f_{n}\}$ such that $m\leq r$, and $T_{3}(f_{i_1}, \delta_{i_1})\cdots(f_{i_m}, \delta_{i_m})$ contains only the row $\bar{\delta}$, it is not difficult to show that
\begin{equation}
\label{equation3}
    N(T_{3}) \leq n^{r} \cdot k^{r}.
\end{equation}
According to the proven above, for any $\bar{\delta}\in E_{k}^{n}$, there exist attributes $f_{i_1},\ldots,$ $f_{i_m}\in \{f_{1},\ldots,f_{n}\}$ such that $m\leq r+1$, and $T_{3}(f_{i_1}, \delta_{i_1})\cdots(f_{i_m}, \delta_{i_m})=T_{3}(f_{1}, \delta_{1})$ $\cdots (f_{n}, \delta_{n})$. Taking into account this equality one can show that
\begin{equation}
    \label{equation4}
    M(T_{3})\leq r+1.
\end{equation}
Using Lemma \ref{lemma_h}, and inequalities (\ref{equation3}) and (\ref{equation4}), we conclude that there exists a deterministic decision tree $\Gamma$ for the table $T_{3}$ with $h(\Gamma) \leq M(T_{3})\log _{2} N(T_{3}) \leq(r+1)^{2} \log _{2} (k n)$. Taking into account that $\psi(f)\leq r$ for any attribute $f\in At(T_{3})$ and the complexity measure $\psi$ has the property (a), we obtain
$$
\psi^{d}(T_{3}) \leq (r+1)^{3} \log _{2} (k n).
$$
Consequently, $\psi^{d}(T_{1}) \leq (r+1)^{3} \log _{2} (k n)$. Taking into account that the complexity measure $\psi$ has the property (c), we obtain $\psi^{i}(T_{1}) \geq n$. Since $T_{1}$ is an arbitrary decision table from $\mathcal{C}$, we have $\operatorname{Dom}^{+}(\mathcal{U}_{\mathcal{C} \psi}^{d i})$ is a finite set. Therefore $\operatorname{typ}(\mathcal{U}_{U \psi}^{d i}) \neq \gamma$. Using Lemma \ref{lemma3} and Corollary \ref{corollary1}, we obtain $\operatorname{typ}(\mathcal{U}_{\mathcal{C} \psi}^{d i}) \in\{\alpha, \beta\}$.
\qed \end{proof}

\begin{proof}[of Proposition \ref{proposition1}]
Let $(\mathcal{C}, \psi)$ be a t-pair. Using Corollary \ref{corollary1}, we conclude that $\operatorname{typ}(\mathcal{U}_{\mathcal{C} \psi}^{i i}) \in\{\alpha, \gamma\}$. Using Corollary \ref{corollary1} and Lemma \ref{lemma3}, we obtain $\operatorname{typ}(\mathcal{U}_{\mathcal{C} \psi}^{d i}) \in\{\alpha, \beta, \gamma\}$. From Lemma \ref{lemma6} it follows that $\operatorname{typ}(\mathcal{U}_{\mathcal{C} \psi}^{a i}) \in\{\alpha, \gamma\}$.

(a) Let $\operatorname{typ}(\mathcal{U}_{\mathcal{C} \psi}^{i i})=\alpha$. Using Lemmas \ref{lemma3} and \ref{lemma4}, we obtain $\operatorname{typ}_{u}(\mathcal{C}, \psi)=t_{1}$.

(b) Let $\operatorname{typ}(\mathcal{U}_{\mathcal{C} \psi}^{i i})=\gamma$ and $\operatorname{typ}(\mathcal{U}_{\mathcal{C} \psi}^{d i})=\alpha$. Using Lemmas \ref{lemma3}, \ref{lemma4}, and \ref{lemma5}, we obtain ${\operatorname{typ}}_{u}(\mathcal{C}, \psi)=t_{2}$.

(c) Let $\operatorname{typ}(\mathcal{U}_{\mathcal{C} \psi}^{i i})=\gamma$ and $\operatorname{typ}(\mathcal{U}_{\mathcal{C} \psi}^{d i})=\beta$. From Lemma \ref{lemma5} it follows that $\operatorname{typ}(\mathcal{U}_{\mathcal{C} \psi}^{i d})=\operatorname{typ}(\mathcal{U}_{\mathcal{C} \psi}^{i a})=\epsilon$. Using Lemmas \ref{lemma3} and \ref{lemma6}, we obtain $\operatorname{typ}(\mathcal{U}_{\mathcal{C} \psi}^{a i})=\alpha$. From this equality and from Lemma \ref{lemma4} it follows that $\operatorname{typ}(\mathcal{U}_{\mathcal{C} \psi}^{a d})=\operatorname{typ}(\mathcal{U}_{\mathcal{C} \psi}^{a a})=$ $\alpha$. Using the equality $\operatorname{typ}(\mathcal{U}_{\mathcal{C} \psi}^{d i})=\beta$, Lemma \ref{lemma3}, and Corollary \ref{corollary1}, we obtain $\operatorname{typ}(\mathcal{U}_{\mathcal{C} \psi}^{d d})=\gamma$. From the equalities $\operatorname{typ}(\mathcal{U}_{\mathcal{C} \psi}^{d d})=\gamma, \operatorname{typ}(\mathcal{U}_{\mathcal{C} \psi}^{a a})=\alpha$ and from Lemmas \ref{lemma2} and \ref{lemma4} it follows that $\operatorname{typ}(\mathcal{U}_{\mathcal{C} \psi}^{d a})=\epsilon$. Thus, $\operatorname{typ}_{u}(\mathcal{C}, \psi)=t_{3}$.

(d) Let $\operatorname{typ}(\mathcal{U}_{\mathcal{C} \psi}^{i i})=\operatorname{typ}(\mathcal{U}_{\mathcal{C} \psi}^{d i})=\gamma$ and $\operatorname{typ}(\mathcal{U}_{\mathcal{C} \psi}^{a i})=\alpha$. Using Lemma \ref{lemma5}, we obtain $\operatorname{typ}(\mathcal{U}_{\mathcal{C} \psi}^{i d})=\operatorname{typ}(\mathcal{U}_{\mathcal{C} \psi}^{i a})=\epsilon$. From Lemma \ref{lemma4} it follows that $\operatorname{typ}(\mathcal{U}_{\mathcal{C} \psi}^{a d})=$ $\operatorname{typ}(\mathcal{U}_{\mathcal{C} \psi}^{a a})=\alpha$. Using Lemma \ref{lemma3} and Corollary \ref{corollary1}, we obtain $\operatorname{typ}(\mathcal{U}_{\mathcal{C} \psi}^{d d})=\gamma$. From this equality, equality $\operatorname{typ}(\mathcal{U}_{\mathcal{C} \psi}^{a a})=\alpha$ and from Lemmas \ref{lemma2} and \ref{lemma4} it follows that $\operatorname{typ}(\mathcal{U}_{\mathcal{C} \psi}^{d a})=\epsilon$. Thus, $\operatorname{typ}_{u}(\mathcal{C}, \psi)=t_{4}$.

(e) Let $\operatorname{typ}(\mathcal{U}_{\mathcal{C} \psi}^{i i})=\operatorname{typ}(\mathcal{U}_{\mathcal{C} \psi}^{d i})=\operatorname{typ}(\mathcal{U}_{\mathcal{C} \psi}^{a i})=\gamma$. Using Lemma \ref{lemma5} we conclude that $\operatorname{typ}(\mathcal{U}_{\mathcal{C} \psi}^{i d})=\operatorname{typ}(\mathcal{U}_{\mathcal{C} \psi}^{i a})=\epsilon$. Using Lemma \ref{lemma3} and Corollary \ref{corollary1}, we obtain $\operatorname{typ}(\mathcal{U}_{\mathcal{C} \psi}^{d d})=\operatorname{typ}(\mathcal{U}_{\mathcal{C} \psi}^{a d})=\operatorname{typ}(\mathcal{U}_{\mathcal{C} \psi}^{a a})=\gamma$. Using Lemma \ref{lemma3}, we obtain $\operatorname{typ}(\mathcal{U}_{\mathcal{C} \psi}^{d a}) \in$ $\{\gamma, \delta, \epsilon\}$. Therefore $\operatorname{typ}_{u}(\mathcal{C}, \psi) \in\{t_{5}, t_{6}, t_{7}\}$.
\qed \end{proof}

\begin{proof}[of Proposition \ref{proposition2}]
Let $(\mathcal{C}, \psi)$ be a limited t-pair. Taking into account that the complexity measure $\psi$ has the property (c), and using Lemma \ref{lemma4}, we obtain $\operatorname{typ}(\mathcal{U}_{\mathcal{C} \psi}^{i i}) \neq \alpha$. Therefore $\operatorname{typ}_{u}(\mathcal{C}, \psi) \neq t_{1}$. Using Lemma \ref{lemma7}, we obtain $\operatorname{typ}_{u}(\mathcal{C}, \psi) \neq t_{4}$. From these relations and Proposition \ref{proposition1} it follows that the statement of the proposition holds.
\qed \end{proof}

\section{Realizable Upper Types of T-Pairs}  \label{S5}

In this section, all realizable upper types of t-pairs are enumerated.
\begin{proposition}
    \label{proposition3.1}
    For any $i \in\{1,2,3,4,5,6,7\}$, there exists a t-pair $(\mathcal{C}, \psi)$ such that
$$
\operatorname{typ}_{u}(\mathcal{C}, \psi)=t_{i} .
$$
\end{proposition}
\begin{proposition}
    \label{proposition3.2}
    For any $i \in\{2,3,5,6,7\}$, there exists a limited $t$-pair $(\mathcal{C}, h)$ such that
$$
\operatorname{typ}_{u}(\mathcal{C}, h)=t_{i} .
$$
\end{proposition}
Proofs of these propositions are based on results obtained for information systems \cite{Moshkov05a}.

Let $U=(A, F)$ be an information system, where attributes from $F$ have values from $E_{k}$, and $\psi$ be a complexity measure over $U$ \cite{Moshkov05a}. Note that $\psi$ is also a complexity measure over the set of decision tables $\mathcal{M}_{k}(F)$. Let $z=(\nu,f_{1},\ldots,f_{n})$ be a problem over $U$. In \cite{Moshkov05a}, three parameters of the problem $z$ were defined: $\psi_{U}^{i}(z)=\psi(f_{1}\cdots f_{n})$ called the complexity of the problem $z$ description, $\psi_{U}^{d}(z)$ -- the minimum complexity of a decision tree with attributes from the set $\{f_{1},\ldots, f_{n}\}$, which solves the problem $z$ deterministically, and $\psi_{U}^{a}(z)$ -- the minimum complexity of a decision tree with attributes from the set $\{f_{1},\ldots, f_{n}\}$, which solves the problem $z$ nondeterministically.

Let $b,c\in \{i, d, a\}$. In \cite{Moshkov05a}, the partial function $\mathcal{U}_{U\psi}^{bc}:\mathbb{N}\rightarrow\mathbb{N}$ was defined:
$$
\mathcal{U}_{U\psi}^{bc}(n)=\max \{\psi_{U}^{b}(z):z\in Probl(U),\psi_{U}^{c}(z)\leq n\} .
$$

The table $\operatorname{typ}_{lu}(U, \psi)$ for the pair $(U, \psi)$ was defined in \cite{Moshkov05a} as follows: this is a table with three rows and three columns in which rows from the top to the bottom and columns from the left to the right are labeled with indices $i,d,a$. The value $\operatorname{typ}(\mathcal{U}_{U\psi}^{bc})$ is in the intersection of the row with the index $b\in \{i,d,a\}$ and the column with the index $c\in \{i,d,a\}$.

We now prove the following proposition:
\begin{proposition}
    \label{proposition3.3}
    Let $U$ be an information system and $\psi$ be a complexity measure over $U$. Then
$$
\operatorname{typ}_{lu}(U, \psi)= \operatorname{typ}_{u}(Tab(U), \psi) .
$$
\end{proposition}

\begin{proof}
    Let $z=(\nu,f_{1},\ldots,f_{n})$ be a problem over $U$ and $T(z)$ be the decision table corresponding to this problem. It is easy to see that $\psi^{i}_{U}(z)=\psi^{i}(T(z))$. One can show the set of decision trees solving the problem $z$ nondeterministically and using only attributes from the set $\{f_{1}, \ldots, f_{n}\}$ (see corresponding definitions in \cite{Moshkov05a}) is equal to the set of nondeterministic decision trees for the table $T(z)$. From here it follows that $\psi_{U}^{a}(z)=\psi^{a}(T(z))$ and $\psi_{U}^{d}(z)=\psi^{d}(T(z))$. Using these equalities, we can show that $\operatorname{typ}_{lu}(U, \psi)= \operatorname{typ}_{u}(Tab(U), \psi)$.
\qed \end{proof}

This proposition allows us to transfer results obtained for information systems in \cite{Moshkov05a} to the case of closed classes of decision tables. Before each of the following seven lemmas, we define a pair $(U, \psi)$, where $U$ is an information system and $\psi$ is a complexity measure over $U$.

Let us define a pair $(U_{1}, \pi)$ as follows: $U_{1}=(\mathbb{N}, F_{1})$, where $F_{1}=\{f\}$ and $f \equiv 0$, and $\pi \equiv 0$.

\begin{lemma}
\label{lemma3.1}
    $\operatorname{typ}_{u}(Tab(U_{1}), \pi)=t_{1}$.
\end{lemma}
\begin{proof}
    From Lemma 4.1 \cite{Moshkov05a} it follows that $\operatorname{typ}_{lu}(U_{1}, \pi)=t_{1}$. Using Proposition \ref{proposition3.3}, we obtain $\operatorname{typ}_{u}(Tab(U_{1}), \pi)=t_{1}$.
\qed \end{proof}

Let us define a pair $(U_{2}, h)$ as follows: $U_{2}=(\mathbb{N}, F_{2})$, where $F_{2}=F_{1}$.

\begin{lemma}
\label{lemma3.2}
    $\operatorname{typ}_{u}(Tab(U_{2}), h)=t_{2}$.
\end{lemma}
\begin{proof}
    From Lemma 4.2 \cite{Moshkov05a} it follows that $\operatorname{typ}_{lu}(U_{2}, h)=t_{2}$. Using Proposition \ref{proposition3.3}, we obtain $\operatorname{typ}_{u}(Tab(U_{2}), h)=t_{2}$.
\qed \end{proof}

Let us define a pair $(U_{3}, h)$ as follows: $U_{3}=(\mathbb{N}, F_{3})$, where $F_{3}=\{l_{i}: i \in \mathbb{N} \setminus\{0\}\}$ and, for any $i \in \mathbb{N} \setminus\{0\}, j \in \mathbb{N}$, if $j \leq i$, then $l_{i}(j)=0$, and if $j>i$, then $l_{i}(j)=1$.

\begin{lemma}
\label{lemma3.3}
    $\operatorname{typ}_{u}(Tab(U_{3}), h)=t_{3}$.
\end{lemma}
\begin{proof}
    From Lemma 4.3 \cite{Moshkov05a} it follows that $\operatorname{typ}_{lu}(U_{3}, h)=t_{3}$. Using Proposition \ref{proposition3.3}, we obtain $\operatorname{typ}_{u}(Tab(U_{3}), h)=t_{3}$.
\qed \end{proof}

Let us define a pair $(U_{4}, \mu)$ as follows: $U_{4}=(\mathbb{N}, F_{4})$, where $F_{4}=F_{3}, \mu(\lambda)=$ $0, \mu(l_{i_{1}} \cdots l_{i_{m}})=1$ if $m=1$ or $m=2$ and $i_{1}>i_{2}, \mu(l_{i_{1}} \cdots l_{i_{m}})=\max \{i_{1}, \ldots, i_{m}\}$ in other cases.

\begin{lemma}
\label{lemma3.4}
    $\operatorname{typ}_{u}(Tab(U_{4}), \mu)=t_{4}$.
\end{lemma}
\begin{proof}
    From Lemma 4.4 \cite{Moshkov05a} it follows that $\operatorname{typ}_{lu}(U_{4}, \mu)=t_{4}$. Using Proposition \ref{proposition3.3}, we obtain $\operatorname{typ}_{u}(Tab(U_{4}), \mu)=t_{4}$.
\qed \end{proof}

Let us define a pair $(U_{5}, h)$ as follows: $U_{5}=(\mathbb{N}, F_{5})$, where $F_{5}=\{f_{i}: i \in$ $\mathbb{N} \setminus\{0\}\}$ and, for any $i \in \mathbb{N} \setminus\{0\}, j \in \mathbb{N}$, if $i=j$, then $f_{i}(j)=1$, and if $i \neq j$, then $f_{i}(j)=0$.

\begin{lemma}
\label{lemma3.5}
    $\operatorname{typ}_{u}(Tab(U_{5}), h)=t_{5}$.
\end{lemma}
\begin{proof}
    From Lemma 4.5 \cite{Moshkov05a} it follows that $\operatorname{typ}_{lu}(U_{5}, h)=t_{5}$. Using Proposition \ref{proposition3.3}, we obtain $\operatorname{typ}_{u}(Tab(U_{5}), h)=t_{5}$.
\qed \end{proof}

Let us define a pair $(U_{6}, h)$ as follows: $U_{6}=(\mathbb{N}, F_{6})$, where $F_{6}=F_{5} \cup G$, $G=\{g_{2 i+1}: i \in \mathbb{N}\}$ and, for any $i \in \mathbb{N}, j \in \mathbb{N}$, if $j \in\{2 i+1,2 i+2\}$, then $g_{2 i+1}(j)=1$, and if $j \notin\{2 i+1,2 i+2\}$, then $g_{2 i+1}(j)=0$.

\begin{lemma}
\label{lemma3.6}
    $\operatorname{typ}_{u}(Tab(U_{6}), h)=t_{6}$.
\end{lemma}
\begin{proof}
    From Lemma 4.6 \cite{Moshkov05a} it follows that $\operatorname{typ}_{lu}(U_{6}, h)=t_{6}$. Using Proposition \ref{proposition3.3}, we obtain $\operatorname{typ}_{u}(Tab(U_{6}), h)=t_{6}$.
\qed \end{proof}

Let us define a pair $(U_{7}, h)$ as follows: $U_{7}=(\mathbb{N}, F_{7})$, where $F_{7}=F_{3} \cup F_{5}$.

\begin{lemma}
\label{lemma3.7}
    $\operatorname{typ}_{u}(Tab(U_{7}), h)=t_{7}$.
\end{lemma}
\begin{proof}
    From Lemma 4.7 \cite{Moshkov05a} it follows that $\operatorname{typ}_{lu}(U_{7}, h)=t_{7}$. Using Proposition \ref{proposition3.3}, we obtain $\operatorname{typ}_{u}(Tab(U_{7}), h)=t_{7}$.
\qed \end{proof}

\begin{proof}[of Proposition \ref{proposition3.1}]
The statement of the proposition follows from Lemmas \ref{lemma3.1}-\ref{lemma3.7}.
\qed \end{proof}

\begin{proof}[of Proposition \ref{proposition3.2}]
The statement of the proposition follows from Lemmas \ref{lemma3.2}, \ref{lemma3.3}, \ref{lemma3.5}, \ref{lemma3.6} and \ref{lemma3.7}.
\qed \end{proof}

\section{Union of T-Pairs}  \label{S6}
In this section we define a union of two t-pairs, which is also a t-pair, and study its upper type.
Let $\tau_{1}=(\mathcal{C}_{1}, \psi_{1})$ and $\tau_{2}=(\mathcal{C}_{2}, \psi_{2})$ be t-pairs, where $\mathcal{C}_{1}\subseteq \mathcal{M}_{k_1}(F_{1})$ and $\mathcal{C}_{2}\subseteq \mathcal{M}_{k_2}(F_{2})$. These two t-pairs will be called \textit{compatible} if $F_{1}\cap F_{2}=\varnothing$ and $\psi_{1}(\lambda)=\psi_{2}(\lambda)$. We now define a t-pair $\tau=(\mathcal{C}, \psi)$, which is called a \textit{union} of compatible t-pairs $\tau_{1}$ and $\tau_{2}$.
\begin{definition} The closed class $\mathcal{C}$ in $\tau$ is defined as follows: $\mathcal{C}=\mathcal{C}_{1}\cup \mathcal{C}_{2}\subseteq \mathcal{M}_{\max(k_{1}, k_{2})} (F_{1}\cup F_{2})$.
    The complexity measure $\psi$ in $\tau$ is defined for any word $\alpha\in (F_{1}\cup F_{2})^{*}$ in the following way:
    if $\alpha\in F_{1}^{*}$, then $\psi(\alpha)=\psi_{1}(\alpha)$,
    if $\alpha\in F_{2}^{*}$, then $\psi(\alpha)=\psi_{2}(\alpha)$,
    if $\alpha$ contains letters from both $F_{1}$ and $F_{2}$, then $\psi(\alpha)$ can have arbitrary value from $\mathbb{N}$. In particular, if $\psi_{1}=\psi_{2}=h$, then as $\psi$  we can use the depth $h$.
\end{definition}
We now consider the upper type of t-pair $\tau=(\mathcal{C}, \psi)$. We denote by $\widetilde{\max}$ the function maximum for the linear order $\alpha\preceq\beta\preceq\gamma\preceq\delta\preceq\epsilon$.
\begin{theorem}
\label{theorem_u}
The equality $\operatorname{typ}(\mathcal{U}_{\mathcal{C} \psi}^{b c})=\widetilde{\max}(\operatorname{typ}(\mathcal{U}_{\mathcal{C}_{1} \psi_{1}}^{b c}), \operatorname{typ}(\mathcal{U}_{\mathcal{C}_{2} \psi_{2}}^{b c}))$ holds for any $b,c\in \{i, d, a\}$ except for the case $bc=da$ and $\operatorname{typ}(\mathcal{U}_{\mathcal{C}_{1} \psi_{1}}^{d a})=\operatorname{typ}(\mathcal{U}_{\mathcal{C}_{2} \psi_{2}}^{d a})=\gamma$. In the last case, $\operatorname{typ}(\mathcal{U}_{\mathcal{C} \psi}^{d a})\in \{\gamma, \delta\}$.
\end{theorem}
\begin{proof}
    Let $n\in \mathbb{N}$ and $b,c\in \{i, d, a\}$. We now define the value $M=\underline{\max}(\mathcal{U}_{1}, \mathcal{U}_{2})$, where $\mathcal{U}_{1}=\mathcal{U}_{\mathcal{C}_{1} \psi_{1}}^{b c}(n)$ and $\mathcal{U}_{2}=\mathcal{U}_{\mathcal{C}_{2} \psi_{2}}^{b c}(n)$. Both $\mathcal{U}_{1}$ and $\mathcal{U}_{2}$ have values from the set $\{\varnothing, \infty\}\cup \mathbb{N}$ (see definitions before Lemma \ref{lemma2}). If $\mathcal{U}_{1}=\mathcal{U}_{2}=\varnothing$, then $M=\varnothing$. If one of $\mathcal{U}_{1}, \mathcal{U}_{2}$ is equal to $\varnothing$ and another one is equal to a number $m\in \mathbb{N}$, then $M=m$. If $\mathcal{U}_{1}, \mathcal{U}_{2}\in \mathbb{N}$, then $M=\max(\mathcal{U}_{1}, \mathcal{U}_{2})$. If at least one of $\mathcal{U}_{1}, \mathcal{U}_{2}$ is equal to $\infty$, then $M=\infty$.

    The following equality follows from the definition of partial function $\mathcal{U}_{\mathcal{C} \psi}^{b c}(n)$, where $n\in \mathbb{N}$ and $b,c\in \{i, d, a\}$:
    $\mathcal{U}_{\mathcal{C} \psi}^{b c}(n)=\underline{\max}(\mathcal{U}_{\mathcal{C}_{1} \psi_{1}}^{b c}(n), \mathcal{U}_{\mathcal{C}_{2} \psi_{2}}^{b c}(n))$. Later in the proof, we will use this equality without special mention.
   From this equality we obtain
   $\operatorname{typ}(\mathcal{U}_{\mathcal{C}_{1} \psi_{1}}^{b c})\preceq \operatorname{typ}(\mathcal{U}_{\mathcal{C} \psi}^{b c})$ and $\operatorname{typ}(\mathcal{U}_{\mathcal{C}_{2} \psi_{2}}^{b c})\preceq \operatorname{typ}(\mathcal{U}_{\mathcal{C} \psi}^{b c})$. We now consider two different cases separately: 1) $\operatorname{typ}(\mathcal{U}_{\mathcal{C}_{1} \psi_{1}}^{b c})=\operatorname{typ}(\mathcal{U}_{\mathcal{C}_{2} \psi_{2}}^{b c})$ and 2) $\operatorname{typ}(\mathcal{U}_{\mathcal{C}_{1} \psi_{1}}^{b c})\neq \operatorname{typ}(\mathcal{U}_{\mathcal{C}_{2} \psi_{2}}^{b c})$.

   1) Let $\operatorname{typ}(\mathcal{U}_{\mathcal{C}_{1} \psi_{1}}^{b c})=\operatorname{typ}(\mathcal{U}_{\mathcal{C}_{2} \psi_{2}}^{b c})$.

   (a) Let $\operatorname{typ}(\mathcal{U}_{\mathcal{C}_{1} \psi_{1}}^{b c})=\operatorname{typ}(\mathcal{U}_{\mathcal{C}_{2} \psi_{2}}^{b c})=\alpha$. Since the functions $\mathcal{U}_{\mathcal{C}_{1} \psi_{1}}^{b c}$ and $\mathcal{U}_{\mathcal{C}_{2} \psi_{2}}^{b c}$ are both bounded from above, we obtain the function $\mathcal{U}_{\mathcal{C} \psi}^{b c}=\underline{\max}(\mathcal{U}_{\mathcal{C}_{1} \psi_{1}}^{b c}, \mathcal{U}_{\mathcal{C}_{2} \psi_{2}}^{b c})$ is also bounded from above. From this, it follows that $\operatorname{typ}(\mathcal{U}_{\mathcal{C} \psi}^{b c})=\widetilde{\max}(\operatorname{typ}(\mathcal{U}_{\mathcal{C}_{1} \psi_{1}}^{b c}),$ $ \operatorname{typ}(\mathcal{U}_{\mathcal{C}_{2} \psi_{2}}^{b c}))=\alpha$. 

   (b) Let $\operatorname{typ}(\mathcal{U}_{\mathcal{C}_{1} \psi_{1}}^{b c})=\operatorname{typ}(\mathcal{U}_{\mathcal{C}_{2} \psi_{2}}^{b c})=\beta$. From the fact that $Dom^{+}(\mathcal{U}_{\mathcal{C}_{1} \psi_{1}}^{b c})$ and $Dom^{+}(\mathcal{U}_{\mathcal{C}_{2} \psi_{2}}^{b c})$ are both finite, we obtain $Dom^{+}(\mathcal{U}_{\mathcal{C} \psi}^{b c})$ is also finite. Similarly, one can show that $\mathcal{U}_{\mathcal{C} \psi}^{b c}$ is unbounded from above on $\mathcal{C}$. From here it follows that $\operatorname{typ}(\mathcal{U}_{\mathcal{C} \psi}^{b c})=\widetilde{\max}(\operatorname{typ}(\mathcal{U}_{\mathcal{C}_{1} \psi_{1}}^{b c}), \operatorname{typ}(\mathcal{U}_{\mathcal{C}_{2} \psi_{2}}^{b c}))=\beta$.

   (c) Let $\operatorname{typ}(\mathcal{U}_{\mathcal{C}_{1} \psi_{1}}^{b c})=\operatorname{typ}(\mathcal{U}_{\mathcal{C}_{2} \psi_{2}}^{b c})=\gamma$. From here, it follows that the function $\psi^{b}$ is unbounded from above on $\mathcal{C}$. From Proposition \ref{proposition1} it follows that $bc$ belongs to the set $\{ii, di, dd, da, ai, ad, aa\}$. Let $c=b$. Using Lemma \ref{lemma4}, we obtain $\operatorname{typ}(\mathcal{U}_{\mathcal{C} \psi}^{b b})=\gamma$. Let $bc \in \{di, ai, ad\}$. Using Lemma \ref{lemma3} and inequalities $\operatorname{typ}(\mathcal{U}_{\mathcal{C}_{1} \psi_{1}}^{b c})\preceq \operatorname{typ}(\mathcal{U}_{\mathcal{C} \psi}^{b c})$ and $\operatorname{typ}(\mathcal{U}_{\mathcal{C}_{2} \psi_{2}}^{b c})\preceq \operatorname{typ}(\mathcal{U}_{\mathcal{C} \psi}^{b c})$, we obtain $\operatorname{typ}(\mathcal{U}_{\mathcal{C} \psi}^{b c})=\gamma$. The only case left is when $bc=da$. Since, there is no $n\in \mathbb{N}$ for which $\mathcal{U}_{\mathcal{C}_{1} \psi_{1}}^{b c}(n)=\infty$ or $\mathcal{U}_{\mathcal{C}_{2} \psi_{2}}^{b c}(n)=\infty$, according to Lemma \ref{lemma2}, we obtain $Dom(\mathcal{U}_{\mathcal{C} \psi}^{b c})$ is an infinite set. Therefore, $\operatorname{typ}(\mathcal{U}_{\mathcal{C} \psi}^{b c})\neq \epsilon$ and hence $\operatorname{typ}(\mathcal{U}_{\mathcal{C} \psi}^{b c})\in \{\gamma, \delta\}$. From Proposition \ref{prop_union} it follows that both cases are possible.

   (d) Let $\operatorname{typ}(\mathcal{U}_{\mathcal{C}_{1} \psi_{1}}^{b c})=\operatorname{typ}(\mathcal{U}_{\mathcal{C}_{2} \psi_{2}}^{b c})=\delta$. From here, it follows that there is no $n\in \mathbb{N}$ for which $\mathcal{U}_{\mathcal{C}_{1} \psi_{1}}^{b c}(n)=\infty$ or $\mathcal{U}_{\mathcal{C}_{2} \psi_{2}}^{b c}(n)=\infty$. Using Lemma \ref{lemma2}, we conclude that $Dom(\mathcal{U}_{\mathcal{C} \psi}^{b c})$ is an infinite set. From the fact that $Dom^{-}(\mathcal{U}_{\mathcal{C}_{1} \psi_{1}}^{b c})$ and $Dom^{-}(\mathcal{U}_{\mathcal{C}_{2} \psi_{2}}^{b c})$ are both finite, we obtain $Dom^{-}(\mathcal{U}_{\mathcal{C} \psi}^{b c})$ is also finite. Therefore, $\operatorname{typ}(\mathcal{U}_{\mathcal{C} \psi}^{b c})=\widetilde{\max}(\operatorname{typ}(\mathcal{U}_{\mathcal{C}_{1} \psi_{1}}^{b c}), \operatorname{typ}(\mathcal{U}_{\mathcal{C}_{2} \psi_{2}}^{b c}))=\delta$.

   (e) Let $\operatorname{typ}(\mathcal{U}_{\mathcal{C}_{1} \psi_{1}}^{b c})=\operatorname{typ}(\mathcal{U}_{\mathcal{C}_{2} \psi_{2}}^{b c})=\epsilon$. Since both $Dom(\mathcal{U}_{\mathcal{C}_{1} \psi_{1}}^{b c})$ and $Dom(\mathcal{U}_{\mathcal{C}_{2} \psi_{2}}^{b c})$ are finite sets, we obtain $Dom(\mathcal{U}_{\mathcal{C} \psi}^{b c})$ is also a finite set. Therefore, $\operatorname{typ}(\mathcal{U}_{\mathcal{C} \psi}^{b c})=\widetilde{\max}(\operatorname{typ}(\mathcal{U}_{\mathcal{C}_{1} \psi_{1}}^{b c}), \operatorname{typ}(\mathcal{U}_{\mathcal{C}_{2} \psi_{2}}^{b c}))=\epsilon$.

   2) Let $\operatorname{typ}(\mathcal{U}_{\mathcal{C}_{1} \psi_{1}}^{b c})\neq \operatorname{typ}(\mathcal{U}_{\mathcal{C}_{2} \psi_{2}}^{b c})$. Denote $f=\mathcal{U}_{\mathcal{C}_{1} \psi_{1}}^{b c}$ and $g=\mathcal{U}_{\mathcal{C}_{2} \psi_{2}}^{b c}$. Let  $\operatorname{typ}(f) \preceq \operatorname{typ}(g)$. We now consider a number of cases. 

   (a) Let $\operatorname{typ}(g)=\epsilon$. From here, it follows that $Dom(g)$ is a finite set. Taking into account this fact, we obtain $Dom(\underline{\max}(f, g))$ is also a finite set. Therefore, $\operatorname{typ}(\underline{\max}(f, g))=\widetilde{\max}(\operatorname{typ}(f), \operatorname{typ}(g))=\epsilon$. Later we will assume that $\operatorname{typ}(g) \neq \epsilon$.

   (b) Let $\operatorname{typ}(f)=\alpha$. Then both $f$ and $g$ are nondecreasing functions, $f$ is bounded from above and $g$ is unbounded from above. From here, it follows that there exists $n_{0}\in \mathbb{N}$ such that $f(n)<g(n)$ for any $n\in \mathbb{N}, n\geq n_{0}$. Using this fact, we conclude that $\underline{\max}(f(n), g(n))=g(n)$ for $n\geq n_{0}$. Therefore, $\operatorname{typ}(\underline{\max}(f, g))=\widetilde{\max}(\operatorname {typ}(f), \operatorname{typ}(g))=\operatorname{typ}(g)$. Later we will assume that $\operatorname{typ}(f) \neq \alpha$. It means we should consider only pairs $(\operatorname{typ}(f), \operatorname{typ}(g))\in \{(\beta, \delta), (\beta, \gamma), (\gamma, \delta)\}$.

   (c) Let $\operatorname{typ}(f)=\beta, \operatorname{typ}(g)=\delta$. From here, it follows that $Dom^{-}(f), Dom^{+}(g)$ are both infinite sets and $Dom^{+}(f), Dom^{-}(g)$ are both finite sets. Taking into account that both $f$ and $g$ are nondecreasing functions, we obtain that there exists $n_{0}\in \mathbb{N}$ such that $f(n)<g(n)$ for any $n\in \mathbb{N}, n\geq n_{0}$. Therefore, $\operatorname{typ}(\underline{\max}(f, g))=\widetilde{\max}(\operatorname{typ}(f), \operatorname{typ}(g))=\operatorname{typ}(g)=\delta$.

   (d) Let $\operatorname{typ}(f)=\beta, \operatorname{typ}(g)=\gamma$. Then $Dom^{+}(\underline{\max}(f, g))$ is an infinite set. Taking into account that $Dom^{-}(g)$ is an infinite set and $Dom^{+}(f)$ is a finite set, we obtain $Dom^{-}(\underline{\max}(f, g))$ is also an infinite set. Therefore, $\operatorname{typ}(\underline{\max}(f, g))=\widetilde{\max}(\operatorname{typ}(f), \operatorname{typ}(g))=\operatorname{typ}(g)=\gamma$.

   (e) Let $\operatorname{typ}(f)=\gamma, \operatorname{typ}(g)=\delta$. From here, it follows that $Dom^{+}(\underline{\max}(f, g))$ is an infinite set and $Dom^{-}(\underline{\max}(f, g))$ is a finite set. Therefore, $\operatorname{typ}(\underline{\max}(f, g))=\widetilde{\max}(\operatorname{typ}(f), \operatorname{typ}(g))=\operatorname{typ}(g)=\delta$.
\qed \end{proof}
The next statement follows immediately from Proposition \ref{proposition1} and Theorem \ref{theorem_u}.
\begin{corollary}
    Let $\tau_{1}$ and $\tau_{2}$ be compatible t-pairs and $\tau$ be a union of these t-pairs. Then the possible values of $\operatorname{typ}_{u}(\tau)$ are in the table shown in Fig. \ref{union_table}, in the intersection of the row labeled with $\operatorname{typ}_{u}(\tau_{1})$ and the column labeled with $\operatorname{typ}_{u}(\tau_{2})$. 
\end{corollary}
\begin{figure}
\begin{minipage}[c]{1.0\textwidth}
\begin{center}
\begin{tabular}{ |c|ccccccc| }
 \hline
  & $t_{1}$ & $t_{2}$ & $t_{3}$ & $t_{4}$ & $t_{5}$ & $t_{6}$ & $t_{7}$\\
  \hline
 $t_{1}$ & $t_{1}$ & $t_{2}$ & $t_{3}$ & $t_{4}$ & $t_{5}$ & $t_{6}$ & $t_{7}$ \\
 $t_{2}$ & $t_{2}$ & $t_{2}$ & $t_{3}$ & $t_{4}$ & $t_{5}$ & $t_{6}$ & $t_{7}$ \\
 $t_{3}$ & $t_{3}$ & $t_{3}$ & $t_{3}$ & $t_{4}$ & $t_{7}$ & $t_{7}$ & $t_{7}$ \\
 $t_{4}$ & $t_{4}$ & $t_{4}$ & $t_{4}$ & $t_{4}$ & $t_{7}$ & $t_{7}$ & $t_{7}$ \\
 $t_{5}$ & $t_{5}$ & $t_{5}$ & $t_{7}$ & $t_{7}$ & $t_{5}, t_{6}$ & $t_{6}$ & $t_{7}$ \\
 $t_{6}$ & $t_{6}$ & $t_{6}$ & $t_{7}$ & $t_{7}$ & $t_{6}$ & $t_{6}$ & $t_{7}$ \\
 $t_{7}$ & $t_{7}$ & $t_{7}$ & $t_{7}$ & $t_{7}$ & $t_{7}$ & $t_{7}$ & $t_{7}$ \\
 \hline
\end{tabular}
\end{center}
\end{minipage}
\caption{ Possible upper types of a union of two compatible t-pairs
}
\label{union_table}
\end{figure}
To finalize the study of unions of t-pairs, we prove the following statement.
\begin{proposition}
\label{prop_union}
    (a) There exist compatible t-pairs $\tau_{1}^{1}$ and $\tau_{2}^{1}$ and their union $\tau^{1}$ such that $\operatorname{typ}_{u}(\tau_{1}^{1})=\operatorname{typ}_{u}(\tau_{2}^{1})=\operatorname{typ}_{u}(\tau^{1})=t_{5}$.

    (b) There exist compatible t-pairs $\tau_{1}^{2}$ and $\tau_{2}^{2}$ and their union $\tau^{2}$ such that $\operatorname{typ}_{u}(\tau_{1}^{2})=\operatorname{typ}_{u}(\tau_{2}^{2})=t_{5}$ and $\operatorname{typ}_{u}(\tau^{2})=t_{6}$.
\end{proposition}
\begin{proof}
    For $i\in \mathbb{N}$, we denote $F_{i}=\{a_{i}, b_{i}, c_{i}\}$ and $G_{i}$ the decision table depicted in Fig.\ref{fig:figE}. We study the t-pair $(\mathcal{T}_{i}, \psi_{i})$, where $\mathcal{T}_{i}$ is the closed class of decision tables from $\mathcal{M}_{2}(F_{i})$, which is equal to $[G_{i}]$, and $\psi_{i}$ is a complexity measure over $\mathcal{M}_{2}(F_{i})$ defined in the following way: $\psi_{i}(\lambda)=0, \psi_{i}(a_{i})=\psi_{i}(b_{i})=\psi_{i}(c_{i})=i$ and $\psi_{i}(\alpha)=i+1$ if $\alpha\in F_{i}^{*}$ and $|\alpha|\geq 2$.

    \begin{figure}
\begin{minipage}[c]{1.0\textwidth}
\begin{center}
$G_{i}=$
\begin{tabular}{ |ccc|c| }
 \hline
 $a_{i}$ & $b_{i}$ & $c_{i}$ &\\
  \hline
 1 & 0 & 0 & $\{1\}$ \\
 0 & 1 & 0 & $\{2\}$ \\
 0 & 0 & 1 & $\{3\}$ \\
 \hline
\end{tabular}
\end{center}
\end{minipage}
    \caption{Decision table $G_{i}$}
    \label{fig:figE}
\end{figure}

    We now study the function $\mathcal{U}_{\mathcal{T}_{i} \psi_{i}}^{d a}$. 
    Since the operations of duplication of columns and permutation of columns do not change the minimum complexity of deterministic and nondeterministic decision trees, we only consider the operations of changing of decisions and removal of columns.

    By these operations, decision tables from $\mathcal{T}_{i}$ can be obtained from $G_{i}$ in three ways: a) only changing of decisions, b) removing one column and changing of decisions, and c) removing two columns and changing of decisions. Figure \ref{fig:figgamma} demonstrates examples of decision tables from $\mathcal{T}_{i}$ for each case. Without loss of generality, we can restrict ourselves to considering these three tables $H_{1}, H_{2},$ and $H_{3}$.

    \begin{figure}
\begin{minipage}[c]{0.3\textwidth}
\begin{center}
$a)$ $H_{1}=$
\begin{tabular}{ |ccc|c| }
 \hline
 $a_{i}$ & $b_{i}$ & $c_{i}$ &\\
  \hline
 1 & 0 & 0 & $d_{1}$ \\
 0 & 1 & 0 & $d_{2}$ \\
 0 & 0 & 1 & $d_{3}$ \\
 \hline
\end{tabular}
\end{center}
\end{minipage}
\begin{minipage}[c]{0.3\textwidth}
\begin{center}
$b)$ $H_{2}=$
\begin{tabular}{ |cc|c| }
 \hline
 $a_{i}$ & $b_{i}$ & \\
  \hline
 0 & 0 & $d_{4}$ \\
 1 & 0 & $d_{5}$ \\
 0 & 1 & $d_{6}$ \\
 \hline
\end{tabular}
\end{center}
\end{minipage}
\begin{minipage}[c]{0.3\textwidth}
\begin{center}
$c)$ $H_{3}=$
\begin{tabular}{ |c|c| }
 \hline
 $c_{i}$ & \\
  \hline
 0 & $d_{7}$ \\
 1 & $d_{8}$ \\
 \hline
\end{tabular}
\end{center}
\end{minipage}
    \caption{Decision tables from closed class $\mathcal{T}_{i}$, where $d_{1},\ldots,d_{8}\in \mathcal{P}(\mathbb{N})$}
    \label{fig:figgamma}
\end{figure}

    (a) There are three different cases for the table $H_{1}$: (i) the sets of decisions $d_{1}, d_{2}, d_{3}$ are pairwise disjoint, (ii) there are $l, t\in \{1, 2, 3\}$ such that $l\neq t, d_{l}\cap d_{t}\neq \varnothing$ and $d_{1}\cap d_{2} \cap d_{3}=\varnothing$, and (iii) $d_{1}\cap d_{2} \cap d_{3}\neq \varnothing$. In the first case, $\psi_{i}^{a}(H_{1})=i$ and $\psi_{i}^{d}(H_{1})=i+1$. In the second case, $\psi_{i}^{a}(H_{1})=i$ and $\psi_{i}^{d}(H_{1})=i$. In the third case, $\psi_{i}^{a}(H_{1})=0$ and $\psi_{i}^{d}(H_{1})=0$.

    (b) There are three different cases for the table $H_{2}$: (i) the sets of decisions $d_{4}, d_{5}, d_{6}$ are pairwise disjoint, (ii) there are $l, t\in \{4, 5, 6\}$ such that $l\neq t, d_{l}\cap d_{t}\neq \varnothing$ and $d_{4}\cap d_{5} \cap d_{6}=\varnothing$, and (iii) $d_{4}\cap d_{5} \cap d_{6}\neq \varnothing$. In the first case, $\psi_{i}^{a}(H_{2})=i+1$ and $\psi_{i}^{d}(H_{2})=i+1$. In the second case, we have either $\psi_{i}^{a}(H_{2})=\psi_{i}^{d}(H_{2})=i+1$ or $\psi_{i}^{a}(H_{2})=\psi_{i}^{d}(H_{2})=i$ depending on the intersecting decision sets. In the third case, $\psi_{i}^{a}(H_{2})=0$ and $\psi_{i}^{d}(H_{2})=0$.

    (c) There are two different cases for the table $H_{3}$: (i) $d_{7}\cap d_{8}=\varnothing$ and (ii) $d_{7}\cap d_{8}\neq \varnothing$. In the first case, $\psi_{i}^{a}(H_{3})=i$ and $\psi_{i}^{d}(H_{3})=i$. In the second case, $\psi_{i}^{a}(H_{3})=0$ and $\psi_{i}^{d}(H_{3})=0$.

    As a result, we obtain that, for any $n\in \mathbb{N}$,
    \begin{equation}
    \label{equ_uda}
        \mathcal{U}_{\mathcal{T}_{i} \psi_{i}}^{d a}(n)=\begin{cases}
    0, & n<i,\\
    i+1, & n\geq i.
    \end{cases}
    \end{equation}

    Let $K$ be an infinite subset of the set $\mathbb{N}$. Denote $F_{K}=\cup_{i\in K} F_{i}$ and $\mathcal{T}_{K} = \cup_{i\in K} [G_{i}]$. It is clear that $\mathcal{T}_{K}$ is a closed class of decision tables from $\mathcal{M}_{2}(F_{K})$. We now define a complexity measure $\psi_{K}$ over $\mathcal{M}_{2}(F_{K})$. Let $\alpha \in F_{K}^{*}$. If $\alpha \in F_{i}^{*}$ for some $i\in K$, then $\psi_{K}(\alpha)=\psi_{i}(\alpha)$. If $\alpha$ contains letters both from $F_{i}$ and $F_{j}$, $i\neq j$, then $\psi_{K}(\alpha)=0$.

    Let $K=\{n_{j}: j\in \mathbb{N}\}$ and $n_{j}<n_{j+1}$ for any $j\in \mathbb{N}$. We define a function $\varphi_{K}:\mathbb{N}\rightarrow \mathbb{N}$ as follows. Let $n\in \mathbb{N}$. If $n<n_{0}$, then $\varphi_{K}(n)=0$. Let, for some $j\in \mathbb{N}$, $n_{j}\leq n<n_{j+1}$. Then $\varphi_{K}(n)=n_{j}$. Using (\ref{equ_uda}), one can show that, for any $n\in \mathbb{N}$,
    $$
    \mathcal{U}_{\mathcal{T}_{K} \psi_{K}}^{d a}(n)=\varphi_{K}(n).
    $$
    Using this equality, one can prove that $\operatorname{typ}(\mathcal{U}_{\mathcal{T}_{K} \psi_{K}}^{d a})=\gamma$ if the set $\mathbb{N}\setminus K$ is infinite and $\operatorname{typ}(\mathcal{U}_{\mathcal{T}_{K} \psi_{K}}^{d a})=\delta$ if the set $\mathbb{N}\setminus K$ is finite.

    Denote $K_{1}^{1}=\{3j:j\in \mathbb{N}\}$, $K_{2}^{1}=\{3j+1:j\in \mathbb{N}\}$ and $K^{1}=K_{1}^{1}\cup K_{2}^{1}$. Denote $\tau_{1}^{1}=(\mathcal{T}_{K_{1}^{1}}, \psi_{K_{1}^{1}})$, $\tau_{2}^{1}=(\mathcal{T}_{K_{2}^{1}}, \psi_{K_{2}^{1}})$ and $\tau^{1}=(\mathcal{T}_{K^{1}}, \psi_{K^{1}})$. One can show that t-pairs $\tau_{1}^{1}$ and $\tau_{2}^{1}$ are compatible and $\tau^{1}$ is a union of $\tau_{1}^{1}$ and $\tau_{2}^{1}$. It is easy to prove that $\operatorname{typ}(\mathcal{U}_{\mathcal{T}_{K_{1}^{1}} \psi_{K_{1}^{1}}}^{d a})=\operatorname{typ}(\mathcal{U}_{\mathcal{T}_{K_{2}^{1}} \psi_{K_{2}^{1}}}^{d a})=\operatorname{typ}(\mathcal{U}_{\mathcal{T}_{K^{1}} \psi_{K^{1}}}^{d a})=\gamma$. Using Proposition \ref{proposition2}, we obtain $\operatorname{typ}_{u}(\tau_{1}^{1})=\operatorname{typ}_{u}(\tau_{2}^{1})=\operatorname{typ}_{u}(\tau^{1})=t_{5}$.

    Denote $K_{1}^{2}=\{2j:j\in \mathbb{N}\}$, $K_{2}^{2}=\{2j+1:j\in \mathbb{N}\}$ and $K^{2}=K_{1}^{2}\cup K_{2}^{2}=\mathbb{N}$. Denote $\tau_{1}^{2}=(\mathcal{T}_{K_{1}^{2}}, \psi_{K_{1}^{2}})$, $\tau_{2}^{2}=(\mathcal{T}_{K_{2}^{2}}, \psi_{K_{2}^{2}})$ and $\tau^{2}=(\mathcal{T}_{K^{2}}, \psi_{K^{2}})$. One can show that t-pairs $\tau_{1}^{2}$ and $\tau_{2}^{2}$ are compatible and $\tau^{2}$ is a union of $\tau_{1}^{2}$ and $\tau_{2}^{2}$. It is easy to prove that $\operatorname{typ}(\mathcal{U}_{\mathcal{T}_{K_{1}^{2}} \psi_{K_{1}^{2}}}^{d a})=\operatorname{typ}(\mathcal{U}_{\mathcal{T}_{K_{2}^{2}} \psi_{K_{2}^{2}}}^{d a})=\gamma$ and $\operatorname{typ}(\mathcal{U}_{\mathcal{T}_{K^{2}} \psi_{K^{2}}}^{d a})=\delta$. Using Proposition \ref{proposition2}, we obtain $\operatorname{typ}_{u}(\tau_{1}^{2})=\operatorname{typ}_{u}(\tau_{2}^{2})=t_{5}$ and $\operatorname{typ}_{u}(\tau^{2})=t_{6}$.
\qed \end{proof}
\section{Proofs of Theorems \ref{theorem1} and \ref{theorem2}}  \label{S7}
First, we consider some auxiliary statements.
\begin{definition}
Let us define a function $\rho:\{\alpha, \beta, \gamma, \delta, \epsilon\} \rightarrow\{\alpha, \beta, \gamma, \delta, \epsilon\}$, as follows:
    $\rho(\alpha)=$ $\epsilon, \rho(\beta)=\delta, \rho(\gamma)=\gamma, \rho(\delta)=\beta, \rho(\epsilon)=\alpha$.
\end{definition}
\begin{proposition}[Proposition 5.1 \cite{Moshkov05a}]
\label{proposition4.1}
Let $X$ be a nonempty set, $f: X \rightarrow \mathbb{N}, g: X \rightarrow \mathbb{N}, \mathcal{U}^{f g}(n)=$ $\max \{f(x): x \in X, g(x) \leq n\}$ and $\mathcal{L}^{g f}(n)=\min \{g(x): x \in X, f(x) \geq n\}$ for any $n \in \mathbb{N}$. Then $\operatorname{typ}(\mathcal{L}^{g f})=\rho(\operatorname{typ}\left(\mathcal{U}^{f g}\right))$.
\end{proposition}

Using Proposition \ref{proposition4.1}, we obtain the following statement.
\begin{proposition}
    \label{proposition5.1}
    Let $(\mathcal{C}, \psi)$ be a t-pair and $b, c \in\{i, d, a\}$. Then $\operatorname{typ}(\mathcal{L}_{\mathcal{C} \psi}^{c b})=$ $\rho(\operatorname{typ}(\mathcal{U}_{\mathcal{C} \psi}^{b c}))$.
\end{proposition}
\begin{corollary}
\label{corollary_last}
    Let $(\mathcal{C}, \psi)$ be a t-pair and $i\in \{1,\ldots,7\}$. Then $\operatorname{typ}_{u}(\mathcal{C}, \psi)=t_{i}$ if and only if $\operatorname{typ}(\mathcal{C}, \psi)=T_{i}$.
\end{corollary}
\begin{proof}[of Theorem \ref{theorem1}]
    The statement of the theorem follows from Propositions \ref{proposition1} and \ref{proposition3.1} and Corollary \ref{corollary_last}.
\qed \end{proof}
\begin{proof}[of Theorem \ref{theorem2}]
    The statement of the theorem follows from Propositions \ref{proposition2} and \ref{proposition3.2} and Corollary \ref{corollary_last}.
\qed \end{proof}

\section{Conclusions }  \label{S8} This paper is devoted to the comparative analysis of
deterministic and nondeterministic decision tree complexity for decision
tables from closed classes. It is a qualitative research: we consider a
finite number of types of the behavior of functions characterizing
relationships among different parameters of decision tables. Future
publications will be related to a quantitative research: we will study lower
and upper bounds on the considered functions.

\section*{Acknowledgements }Research reported in this publication was
supported by King Abdullah University of Science and Technology (KAUST).

\bibliographystyle{splncs04}
\bibliography{closedclasses}

\begin{thebibliography}{10}
\providecommand{\url}[1]{\texttt{#1}}
\providecommand{\urlprefix}{URL }
\providecommand{\doi}[1]{https://doi.org/#1}

\bibitem{AbouEishaACHM19}
AbouEisha, H., Amin, T., Chikalov, I., Hussain, S., Moshkov, M.: Extensions of
  Dynamic Programming for Combinatorial Optimization and Data Mining,
  Intelligent Systems Reference Library, vol.~146. Springer (2019)

\bibitem{book20}
Alsolami, F., Azad, M., Chikalov, I., Moshkov, M.: Decision and Inhibitory
  Trees and Rules for Decision Tables with Many-valued Decisions, Intelligent
  Systems Reference Library, vol.~156. Springer (2020)

\bibitem{Atminas17}
Atminas, A., Lozin, V.V., Moshkov, M.: {WQO} is decidable for factorial
  languages. Inf. Comput.  \textbf{256},  321--333 (2017)

\bibitem{BlumI87}
Blum, M., Impagliazzo, R.: Generic oracles and oracle classes (extended
  abstract). In: 28th Annual Symposium on Foundations of Computer Science, Los
  Angeles, California, USA, 27-29 October 1987. pp. 118--126. {IEEE} Computer
  Society (1987)

\bibitem{BorosHIK97}
Boros, E., Hammer, P.L., Ibaraki, T., Kogan, A.: Logical analysis of numerical
  data. Math. Program.  \textbf{79},  163--190 (1997)

\bibitem{BorosHIKMM00}
Boros, E., Hammer, P.L., Ibaraki, T., Kogan, A., Mayoraz, E., Muchnik, I.B.: An
  implementation of logical analysis of data. {IEEE} Trans. Knowl. Data Eng.
  \textbf{12}(2),  292--306 (2000)

\bibitem{Boutell04}
Boutell, M.R., Luo, J., Shen, X., Brown, C.M.: Learning multi-label scene
  classification. Pattern Recognit.  \textbf{37}(9),  1757--1771 (2004)

\bibitem{BreimanFOS84}
Breiman, L., Friedman, J.H., Olshen, R.A., Stone, C.J.: {Classification and
  Regression Trees}. Wadsworth and Brooks (1984)

\bibitem{BuhrmanW02}
Buhrman, H., de~Wolf, R.: Complexity measures and decision tree complexity: a
  survey. Theor. Comput. Sci.  \textbf{288}(1),  21--43 (2002)

\bibitem{ChikalovLLMNSZ13}
Chikalov, I., Lozin, V.V., Lozina, I., Moshkov, M., Nguyen, H.S., Skowron, A.,
  Zielosko, B.: Three Approaches to Data Analysis - Test Theory, Rough Sets and
  Logical Analysis of Data, Intelligent Systems Reference Library, vol.~41.
  Springer (2013)

\bibitem{FurnkranzGL12}
F{\"{u}}rnkranz, J., Gamberger, D., Lavrac, N.: Foundations of Rule Learning.
  Cognitive Technologies, Springer (2012)

\bibitem{HartmanisH87}
Hartmanis, J., Hemachandra, L.A.: One-way functions, robustness, and the
  non-isomorphism of {NP}-complete sets. In: Proceedings of the Second Annual
  Conference on Structure in Complexity Theory, Cornell University, Ithaca, New
  York, USA, June 16-19, 1987. {IEEE} Computer Society (1987)

\bibitem{Lozin21}
Lozin, V.V., Moshkov, M.: Critical properties and complexity measures of
  read-once {B}oolean functions. Ann. Math. Artif. Intell.  \textbf{89}(5-6),
  595--614 (2021)

\bibitem{Molnar22}
Molnar, C.: Interpretable Machine Learning. A Guide for Making Black Box Models
  Explainable. 2 edn. (2022), \url{christophm.github.io/interpretable-ml-book/}

\bibitem{Moshkov89}
Moshkov, M.: On depth of conditional tests for tables from closed classes. In:
  Markov, A.A. (ed.) Combinatorial-Algebraic and Probabilistic Methods of
  Discrete Analysis (in Russian), pp. 78--86. Gorky University Press, Gorky
  (1989)

\bibitem{Moshkov95}
Moshkov, M.: About the depth of decision trees computing {B}oolean functions.
  Fundam. Informaticae  \textbf{22}(3),  203--215 (1995)

\bibitem{Moshkov05a}
Moshkov, M.: Comparative analysis of deterministic and nondeterministic
  decision tree complexity. {L}ocal approach. Trans. Rough Sets  \textbf{4},
  125--143 (2005)

\bibitem{Moshkov05}
Moshkov, M.: Time complexity of decision trees. Trans. Rough Sets  \textbf{3},
  244--459 (2005)

\bibitem{Moshkov20}
Moshkov, M.: Comparative Analysis of Deterministic and Nondeterministic
  Decision Trees, Intelligent Systems Reference Library, vol.~179. Springer
  (2020)

\bibitem{MPZ08}
Moshkov, M., Piliszczuk, M., Zielosko, B.: Partial Covers, Reducts and Decision
  Rules in Rough Sets - Theory and Applications, Studies in Computational
  Intelligence, vol.~145. Springer (2008)

\bibitem{MoshkovZ11}
Moshkov, M., Zielosko, B.: Combinatorial Machine Learning - {A} Rough Set
  Approach, Studies in Computational Intelligence, vol.~360. Springer (2011)

\bibitem{Pawlak81}
Pawlak, Z.: Information systems theoretical foundations. Inf. Syst.
  \textbf{6}(3),  205--218 (1981)

\bibitem{Pawlak91}
Pawlak, Z.: Rough Sets - Theoretical Aspects of Reasoning about Data, Theory
  and Decision Library: Series {D}, vol.~9. Kluwer (1991)

\bibitem{PawlakS07}
Pawlak, Z., Skowron, A.: Rudiments of rough sets. Inf. Sci.  \textbf{177}(1),
  3--27 (2007)

\bibitem{Post41}
Post, E.: Two-valued Iterative Systems of Mathematical Logic, Annals of Math.
  Studies, vol.~5. Princeton Univ. Press, Princeton-London (1941)

\bibitem{Quinlan93}
Quinlan, J.R.: {C4.5:} Programs for Machine Learning. Morgan Kaufmann (1993)

\bibitem{Robertson04}
Robertson, N., Seymour, P.D.: Graph minors. {XX.} {W}agner's conjecture. J.
  Comb. Theory, Ser. {B}  \textbf{92}(2),  325--357 (2004)

\bibitem{RokachM07}
Rokach, L., Maimon, O.: Data Mining with Decision Trees - Theory and
  Applications, Series in Machine Perception and Artificial Intelligence,
  vol.~69. World Scientific (2007)

\bibitem{Tardos89}
Tardos, G.: Query complexity, or why is it difficult to separate ${NP}^{A}\cap
  co{NP}^{A}$ from ${P}^{A}$ by random oracles ${A}$? Comb.  \textbf{9}(4),
  385--392 (1989)

\bibitem{Vens08}
Vens, C., Struyf, J., Schietgat, L., Dzeroski, S., Blockeel, H.: Decision trees
  for hierarchical multi-label classification. Mach. Learn.  \textbf{73}(2),
  185--214 (2008)

\bibitem{Zhou12}
Zhou, Z., Zhang, M., Huang, S., Li, Y.: Multi-instance multi-label learning.
  Artif. Intell.  \textbf{176}(1),  2291--2320 (2012)

\end{thebibliography}
\end{document}